# Only what exists can cause:
# An intrinsic powers view of free will

Giulio Tononi,[1]* Larissa Albantakis,[1] Melanie Boly,[1] Chiara Cirelli,[1] Matteo Grasso,[1] Christof Koch[2]

[1]University of Wisconsin–Madison, Department of Psychiatry, Madison (WI), USA; [2]Allen Institute for Brain Science, Seattle (WA), USA; *Correspondence to gtononi@wisc.edu

This essay addresses the implications of integrated information theory (IIT) for free will. IIT is a theory of what consciousness is and of how its presence and quality can be accounted for in physical terms. According to IIT, the presence of consciousness is accounted for by a maximum of cause–effect power in the brain. Moreover, the way an experience feels is accounted for by how that cause–effect power is structured. If IIT is right, we do have free will in a genuine sense: we have alternatives, reasons, and values, we make decisions, and we—not our neurons or atoms—are the cause of our willed actions and bear responsibility for them. IIT's argument for genuine free will hinges on the proper understanding of consciousness as intrinsic existence, captured by its intrinsic powers ontology: what exists absolutely, in physical terms, are intrinsic entities, and only what exists can cause.[i]

---

[i] Sections 5–8 draw substantially on a chapter on free will from a book in preparation (subject to the associated licensing restrictions): Tononi, G. (in preparation). *On Being: Ontological and metaphysical implications of integrated information theory (IIT).*

It matters not how straight the gate,
How charged with punishments the scroll,
I am the master of my fate:
I am the captain of my soul.
*William Ernest Henley*

# 1 Introduction

What could be easier than raising my hand if I decide to do so? Yet figuring out what it means to decide among alternatives, to will an action, and to will it freely is not a simple enterprise. Many neuroscientists, psychologists, and philosophers hold that we cannot possibly have genuine free will—free will *in a genuine sense* (sometimes called "metaphysical" free will). This is because in the end, in a causally closed universe, all our actions must be determined by our neurons (and ultimately by our atoms), if they are determined at all and not random. Therefore, they claim, metaphysical free will is either an illusion or an incoherent notion. We must be content with a practical or social notion of freedom—the ability to decide and carry out an action voluntarily and autonomously, rather than under duress or, say, under the influence of alcohol.

This essay argues otherwise. I can have free will in a genuine sense: I can have genuine alternatives in the sense that I can choose among them, that I can weigh reasons to choose among them, that I can decide and intend my actions, that I can cause, control, and execute those actions, and I therefore bear responsibility. The argument on which these conclusions rest derives from integrated information theory (IIT) (Albantakis et al., 2023; Tononi et al., 2016) and requires coming to grips with what consciousness is and with the ontology that follows from it. It also requires a worked-out notion of what causes what. This essay aims to present the argument concisely, which comes at a price. We will have to forgo a full explanation of key concepts and only briefly engage with the vast literature on free will and its compatibility with determinism (Dennett, 1984; Harris, 2012; Kane, 2002, 2005, 2011; List, 2019; Maoz & Sinnott-Armstrong, 2022; McKenna & Pereboom, 2016; Mele, 2014; O'Connor, 2000; Pereboom, 2009; Schurger & Uithol, 2015; Watson, 2003). The essay will also not examine the empirical evidence about illusions of free will (Wegner, 2002) and about the ability to predict a subject's decisions based on brain activity (Haggard, 2019; Libet et al., 1983). Nor will it address the cultural, historical, and legal aspects of free will. However, if the thesis that we have genuine free will is correct, it will have many implications for all those aspects.

After introducing a simple free will scenario (**sec. 2**), the essay opens with an outline of IIT's ontology (**sec. 3**), complemented by an account of actual causation (**sec. 4**). A brief overview of the requirements for free will (**sec. 5**) is followed by the implications of IIT for genuine free will (**sec. 6**). IIT's intrinsic powers ontology is then contrasted with standard substrate ontologies (**sec. 7**). The essay continues with a discussion of indeterminism and the implications of IIT for freedom, self-changing actions, and responsibility (**sec. 8**). A survey of IIT's view of free will in relation to major philosophical positions follows (**sec. 9**). The essay ends with a brief consideration of experimental tests (**sec. 10**).

# 2 A simple free will scenario

To make matters explicit, we will consider a simple but paradigmatic free will scenario. **Fig. 1** shows four successive experiences I might go through: (1) imagining alternative courses of action, (2) envisioning one or more reasons favoring one action or another, (3) coming to a decision, and finally (4) controlling an action that carries out the intended outcome. Later, the essay will argue that each is an essential ingredient of free will. Of course, realistic scenarios encompass many variants. For example, I may go back and forth among alternatives and reasons, change my mind, sleep over a difficult decision before taking it, and so on. Other ingredients are also important. For example, the set of beliefs that can influence a decision may be small or large, shallow or deep; they may involve notions of the self and autobiographical memories and include innate or acquired values. As we will see, the broader one's understanding of the context of a decision, the more free will can be exercised. As we will also see, freely willed decisions can change who we become as well as the world around us, all of which progressively increases our responsibilities.

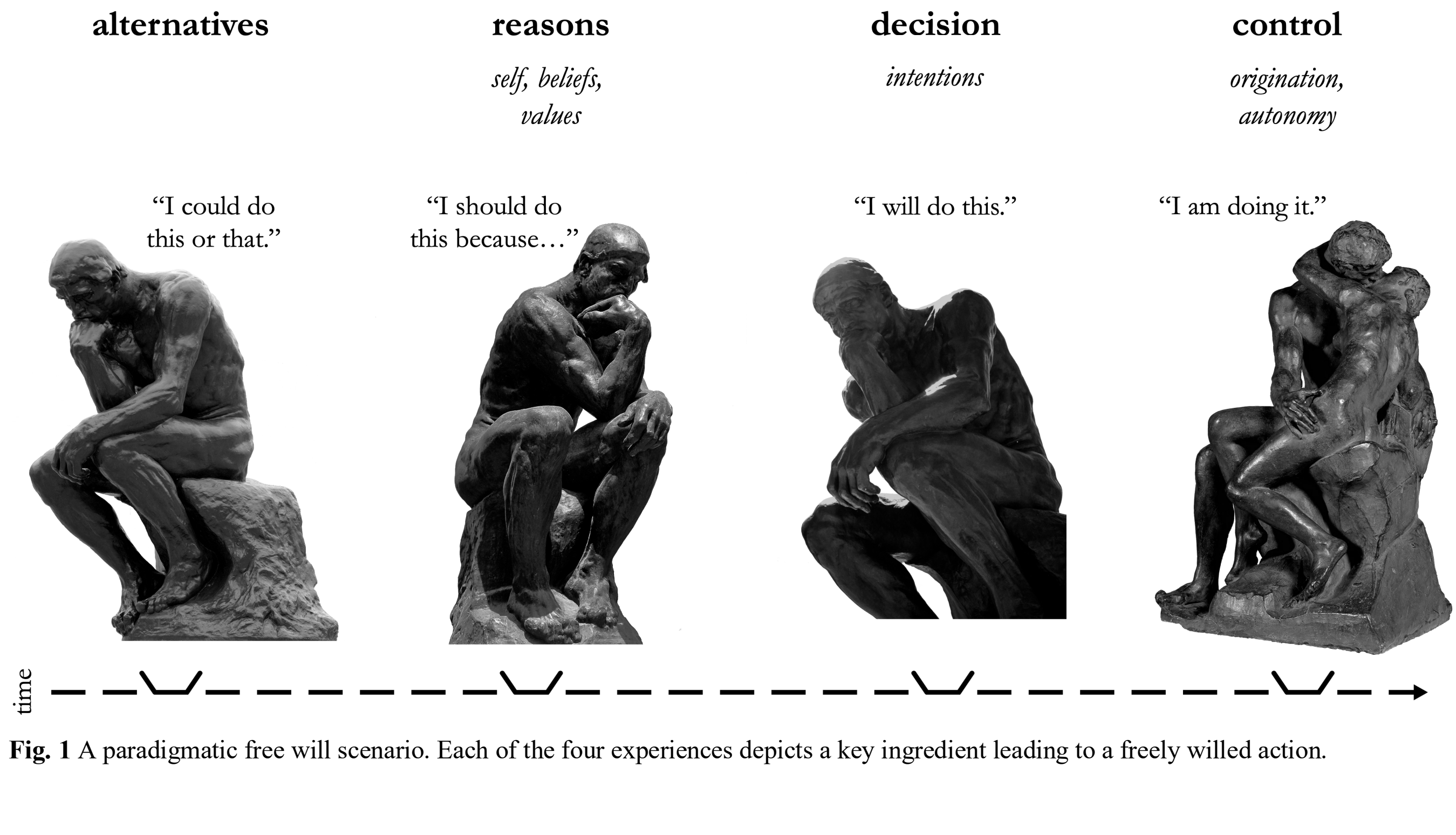

**Fig. 1** A paradigmatic free will scenario. Each of the four experiences depicts a key ingredient leading to a freely willed action.

# 3 Ontology: an experience as a $\Phi$-structure unfolded from a complex

The critical starting point is a proper ontology: a principled, coherent notion of what exists when I, as a conscious subject, consider alternatives, reason, decide, and execute an action. Before concluding whether free will can exist or not, we need a theory of what consciousness is and what it takes to have it. This requirement is not exclusive to free will—for example, meaning, too, can only be understood on the basis of a theory of consciousness. But it is especially acute in the case of free will because, as we will argue, only what exists can cause.

This paper relies on the ontology of consciousness proposed by IIT. If IIT is correct, as can be established through empirical tests, it has critical consequences for free will. The bare bones of IIT will be sketched below, and only to the extent that they are required for the argument. A proper introduction can be found elsewhere (Albantakis et al., 2023; Hendren et al., 2024 [IIT Wiki]; Tononi et al., 2016).

## 3.1 Phenomenal and physical existence

IIT addresses the problem of consciousness starting from phenomenology—the existence of my own experience, which is immediate and irrefutable—rather than from the behavioral, functional, or neural correlates of experience.

The first step is to acknowledge that the experience I am having, here and now, *exists*. When I come into being, say, awakening from a dreamless slumber, my very experience demonstrates that there is something, not nothing. *Phenomenal existence* is IIT's *0th axiom* because there is no more fundamental and certain starting point. As recognized by Descartes and others, the existence of experience is given immediately—it is not something I infer from other evidence. And it is irrefutable, because even doubting is an experience, thus confirming the validity of the axiom.

It is critical to recognize the ontological primacy of phenomenal existence, but it does not lead very far by itself. There are countless regularities in my experiences, and they cry out for a good explanation. For example, I am confident, here and now, that the scene I am experiencing is the same as the one I remember experiencing before I closed and reopened my eyes. I am convinced, here and now, that my hands remain attached to my body no matter how I move them, and so on. To account for these feelings of regularity and invariance, we make countless *inferences to a good explanation*, whether we realize it or not. The most fundamental of these are codified in IIT as realism, physicalism, and atomism.

*Realism* assumes that there are "things" that exist independent of my experiencing them, and they persist even when, for example, I fall asleep and cannot interact with them. *Physicalism* assumes an *operational* criterion for assessing the existence of things outside of my own experience: I can be confident that something exists—whether it is my hand, a rock, a distant star, or an elementary particle—if I can show or infer that it has cause–effect power, in the sense that it can "take and make a difference," and it does so in a way that is reliable and persisting. *Atomism* (operational reductionism) assumes that ideally, to leave nothing out, an explanation should start from the smallest units that can take and make a difference.

Based on these assumptions, IIT introduces its *0th postulate* or "principle of being": to exist in physical terms means to have cause–effect power—being able to take and make a difference.[ii] In other words, physical existence is defined purely operationally, from the perspective of a conscious observer, with no residual "intrinsic" properties (such as mass or charge).[iii] Furthermore, physical existence should be conceived of as cause–effect power all the way down—namely down to the finest, "atomic" units that can take and make a difference.[iv] For a finite, discrete, binary system one would ideally obtain an atomic transition probability matrix (TPM): a matrix that specifies the probability of transitioning for every system state into every system state. The TPM will also reflect how each unit responds to manipulations of the state of any other set of units.

From these foundations, IIT accounts for consciousness in the following way. We first use introspection and reasoning to identify the essential properties of consciousness—the axioms of phenomenal existence. We then consider what could

[ii] The principle of being is related to the Eleatic principle expressed in Plato's *Sophist*. The Eleatic principle says that for something to exist, it must take *or* make a difference, whereas the principle of being says that to exist something must take *and* make a difference, as well as provide itself with a difference.

[iii] This common use of the term "intrinsic," referring to a property that "inheres" to a substrate, is different from the use in IIT, where intrinsic means "for itself," "from within." According to IIT, what one can assess operationally are ultimately just conditional probabilities ("if I do this, I see that.") This also implies that all "categorical" properties (what something is like) are ultimately "dispositional" (what powers something has).

[iv] Fundamental physics is open about ultimate constituents (particles as opposed to fields) and about the discrete or continuous nature of the universe. From first principles (operational cause–effect power), IIT assumes discrete micro units, micro updates, and micro states.

account for each axiom in terms of cause–effect power; that is, we formulate an essential phenomenal property as an essential property of the substrate of consciousness— yielding the postulates of physical existence. In this way, we obtain a set of criteria that allow us to identify and characterize a substrate of consciousness (say, a set of cortical neurons). These steps are summarized in brief below and in **fig. 2**.

### 3.2 Axioms: the essential properties of phenomenal existence

Using introspection and reasoning, one can identify five *essential* properties of consciousness—namely, intrinsicality, information, integration, exclusion, and composition. These are known as the *axioms* of phenomenal existence because they are immediate and irrefutably true of every conceivable experience. They are immediate in the sense that they are evident without requiring any inference, and they are irrefutable in the sense that it is impossible or absurd to conceive of an experience that lacked them.

**Intrinsicality** means that experience is *intrinsic*: it exists *for itself*. Thus, when I experience the scene in front of me, the scene is for me,[v] experienced from within, from my intrinsic perspective. This property is true of every conceivable experience: if I conceive of an experience that would be for someone else, it would then be theirs, confirming the axiom; conversely, an experience that is for no one is inconceivable. This reminds us that intrinsic existence—consciousness—is the only existence worth being. We can also say that experience, as intrinsic existence, is *absolute* existence, rather than existence relative to someone else.

**Information** means that experience is *specific*: it is *this one*. Thus, when I see my bedroom, the experience I am having is this one, not another one. Again, this property is true of every conceivable experience: if I conceive of an experience that would be another one, it would then be that one, confirming the axiom; conversely, an experience that is not this or any other one, but generic, is inconceivable. Because an experience is always this one, it also differs from a large repertoire of possible experiences, a property called *differentiation*.

**Integration** means that experience is *unitary*: it is *a whole*, irreducible to separate experiences. Thus, in front of me I see both the left and the right side of my bedroom, and what I see cannot be reduced to a left side and a right side that are experienced separately. Integration is true of every conceivable experience: if I conceive of an experience containing seemingly independent parts, like a ‘flash’ and a ‘bang,’ the experience would then be a ‘flashbang,’ which is a whole, confirming the axiom; conversely, an experience such that the left side would be experienced by me and the right side by someone else is inconceivable.

**Exclusion** means that experience is *definite*: it is *this whole*, containing all it contains, neither less nor more. For example, my visual experience has a border: it includes all the visual field—its left and right side. It excludes my experiencing less—say, the left side only but not the right side—and my experiencing more—say, a periphery that extends to the back of my head. Exclusion is also true of every conceivable experience: if I conceive of an experience containing less or more, it would then be that whole, confirming the axiom; conversely, an experience that is not all it is, but indefinite, is inconceivable.

**Composition** means that experience is *structured*: it is *the way it is*, being composed of phenomenal distinctions and the relations that bind them, yielding a phenomenal structure that feels the way it feels. For instance, I can distinguish a body, a hand, and a book; the hand is attached to the body and lying on the book. Again, composition is true of every conceivable experience: if I conceive of an experience that might be some other way, it would then be structured that way, confirming the axiom. Conversely, an experience that is not this way or that way, but no way, is inconceivable.[vi]

### 3.3 Postulates: the essential properties of physical existence

If every experience is characterized by the five essential properties of phenomenal existence, the substrate of consciousness should be characterized by the same properties in terms of physical existence, understood operationally as cause–effect power established by observing and manipulating its constituting units. IIT thus formulates the five phenomenal axioms as five physical *postulates*:

**Intrinsicality** means that the cause–effect power of a substrate of consciousness must be *intrinsic*: it must take and make a difference *within itself*. The units of a complex are called *intrinsic units*. Every unit outside the substrate of consciousness is considered a *background condition*.

**Information** means that its cause–effect power must be *specific*: it must be in *this state* and select *this cause–effect state*. The selected state is the one that maximizes intrinsic information (*ii*).[vii]

**Integration** means that its cause–effect power must be *unitary*: it must specify its cause–effect state as *a whole set* of units, irreducible to separate subsets. Substrate units, too, must be irreducible. Irreducibility is measured by integrated information ($\varphi_s$) over a substrate's minimal partition, which serves as a quantifier of integrated existence (how much something exists as *one* thing).

[v] Here, “me” should be understood simply as the subject of experience, not as a separate entity that does the experiencing, even less as a self in a conceptual or autobiographical sense.

[vi] States of “pure consciousness,” “pure presence,” or “naked awareness” are sometimes described as “contentless” because subjects who can achieve them experience no objects, no self, and no thoughts. However, subjects do experience something akin to a vast expanse, often described as “luminous,” which is indeed a content, structured in a specific way.

[vii] This is a measure of intrinsic difference that can be shown to capture uniquely the postulates of existence, intrinsicality, and information ((Barbosa et al., 2020). A key property of this measures is that it is sensitive to a tension between expansion and dilution.

**Exclusion** means that its cause–effect power must be *definite*: it must specify its cause–effect state as *this whole set* of units, all of them, neither less nor more. This is the set of units that is maximally irreducible, as measured by maximum $\varphi_s$ ($\varphi_s$*). This set is called a maximal substrate, or *complex*. Note that the complex's units, called *intrinsic units*, may be macro units that update over macro updates, as long as they satisfy the

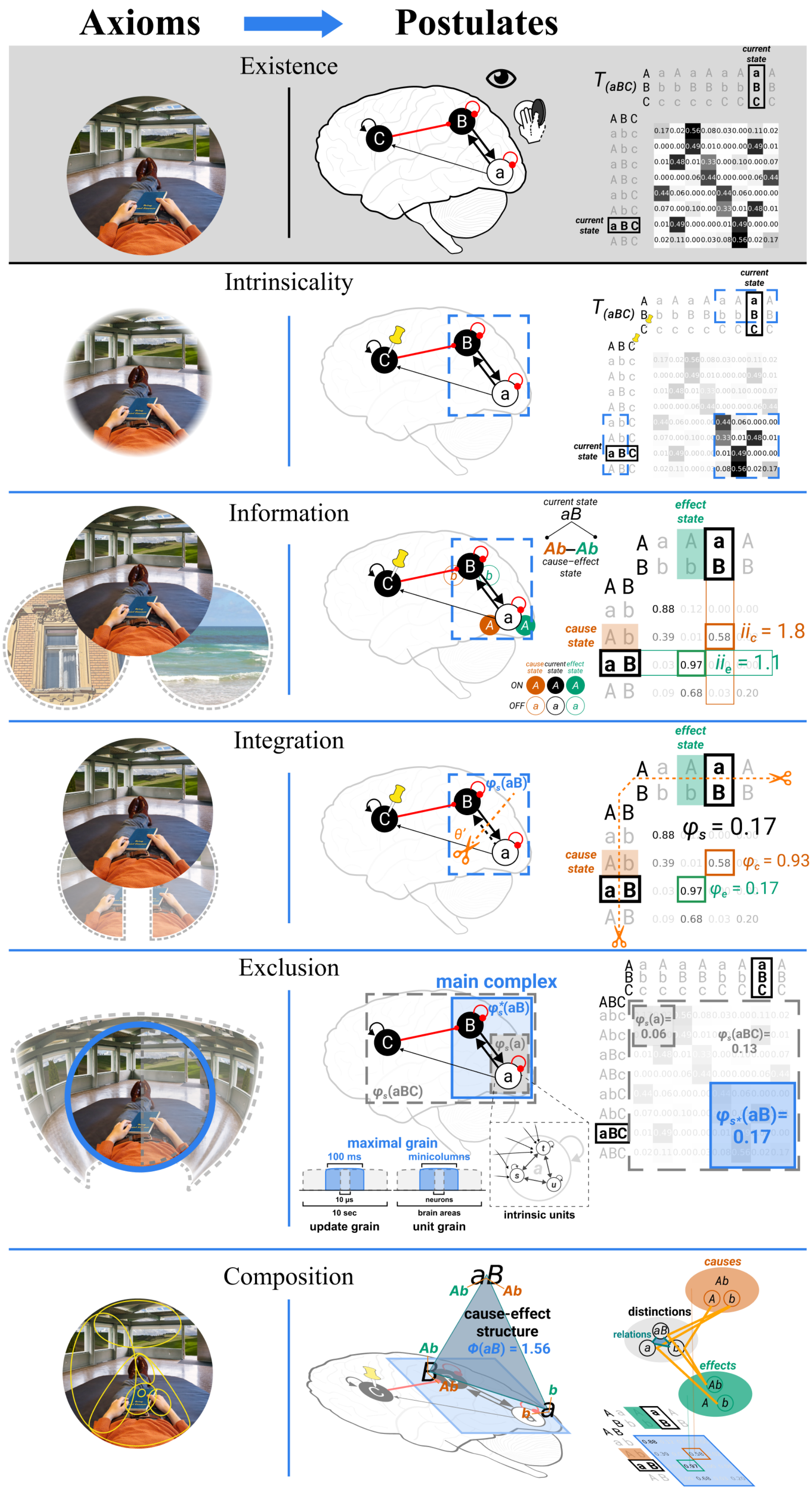

**Fig. 2** IIT's axioms of phenomenal existence and postulates of physical existence. My experience exists, immediately and indubitably. Based on its regularities, I infer the existence of a world independent of me. I call it "physical" if I can operationally observe (eye) and manipulate (hand) the states (uppercase ON, lowercase OFF) of a substrate (e.g., a brain) whose units can "take and make a difference." This yields conditional probabilities (right) that capture its cause–effect power, summarized by a substrate model (left). Just as my experience exists for itself, its substrate must have cause–effect power for itself. The dotted blue line on the substrate (left) indicates a candidate substrate (aB) whose causal powers may be analyzed within itself as a transition probability matrix (TPM, right), with background conditions (C) fixed (pin). Just as my experience is specific, its substrate, in its current state (aB), must select a specific cause-effect state (s′ = {Ab, Ab}, cause in red, effect in green). This is the state that maximizes intrinsic information (ii) for the substrate as a whole. And just as my experience, being specific, is different from countless other experiences, so is the cause–effect state specified by other states of the substrate (e.g., Ab and AB). Just as my experience is unitary, the cause–effect power of its substrate must be unitary. If so, partitioning the substrate (e.g., unidirectionally between unit a and unit B, left) must reduce the intrinsic information it specifies (shown on the TPM, right). The irreducibility of the cause-effect state specified by the substrate as a whole to that specified by separate parts—how much a substrate exists as one substrate—is measured by integrated information ($\varphi s$). Just as my experience is definite, the cause–effect power of its substrate must be definite. By the principle of maximal existence, the substrate of consciousness (the main complex, here aB, indicated by the blue outline) is the one for which integrated information is maximal ($\varphi s^*$) and whose intrinsic units are maximally irreducible within, thus excluding all overlapping sets (e.g., a and aBC) and all overlapping unit and update a grains (e.g. micro units s, t, and u with shorter updates. Just as my experience is structured by distinctions and relations, the cause–effect power of the complex must be structured by distinctions and relations specified by subsets of units over subsets of units (right). Distinctions (black) link a cause (red) and an effect (green) through a mechanism. Relations (shades of blue) bind distinctions whose causes and/or effects overlap. Together, distinctions and relations compose the cause-effect structure "unfolded" from a complex (left). The sum of their $\varphi$ values corresponds to the complex's structure integrated information ($\Phi$).

postulates, among them that they are maximally irreducible.[viii]

**Composition** means that its cause–effect power must be *structured*: subsets of units must specify cause–effects over subsets of units (*distinctions*) that can overlap with one another (*relations*), yielding a *cause–effect structure*, or Φ-*structure*, that is *the way it is*.

A substrate of consciousness—say, a set of neurons in the cerebral cortex—must comply with all postulates. In principle, we can check this by obtaining a complete TPM of the system—an overview of how its units respond to all possible perturbations of their state, given certain background conditions. From the TPM, we can determine which set of units constitutes a complex and *unfold* all its causal powers—its *Φ*-structure. A complex specifying a *Φ*-structure satisfies, in physical terms, all the essential properties of phenomenal existence and should be considered an *intrinsic entity*—one that exists intrinsically, for itself. And just as experience exists absolutely—as we know immediately and indubitably—an intrinsic entity should also be considered as existing absolutely, not relative to something else.

### 3.4 The explanatory identity: An experience as the Φ-structure unfolded from a complex

On this basis, IIT asserts an *explanatory identity*: an experience is identical to the *Φ*-structure unfolded from a complex. In other words, all phenomenal properties of an experience—its quality or how it feels—are accounted for in full by the properties of the cause–effect structure unfolded from a maximal substrate, with no additional ingredients (**fig. 3**). Thus, all the contents of an experience here and now—including spatial extendedness, temporal flow, objects, colors and sounds, thoughts, intentions, decisions, and beliefs, doubts and convictions, hopes and fears, memories and expectations—correspond to *sub-structures* in a cause–effect structure (*Φ-folds* in a *Φ*-structure). Moreover, the sum total of the $\varphi$ values[ix] of the distinctions and relations that compose the *Φ*-structure measures its *structure integrated information Φ* ("big PHI") and corresponds to the quantity of consciousness—the content of an experience. And just as a *Φ*-structure, corresponding to an experience as a whole, can be said to exist absolutely, the sub-structures within it, corresponding to contents of the experience, can be said to exist absolutely within the experience.[x]

The validity of IIT as an account of consciousness can be assessed by considering its ability to explain and predict both the presence of consciousness and its content in the one case we are sure of, our own experience and our own brain. Regarding the presence, we can assess whether brain areas thought to contribute to the neural substrate of consciousness comply with the postulates, whereas other areas do not. For example, posterior-central cortical areas are organized as a set of grids stacked upon one another in series and in parallel, linked

[viii] They must be maximally irreducible within, as they are not complexes themselves, and the complement of macro units is the one that maximizes the irreducibility of the complex. Also, because they are not complexes, composition does not apply to units. The unit grain is the relevant scale for causal observations and manipulations, whether they be proteins, synapses, dendrites, or other sub-cellular compartments; neurons, groups of neurons, or columns; and so on. The update grain is the relevant timescale of the causal operations, whether tens, hundreds, or thousands of milliseconds, and the optimal mapping of micro states into two unit states (ON or OFF).

[ix] The irreducibility of the distinctions ($\varphi_d$) and relations ($\varphi_r$) that compose the *Φ*-structure measures how much each of them exists *within* the experience.

[x] In fact, according to IIT, even contents such as abstract concepts (truth and beauty, values and norms, numbers, and propositions), only exist as sub-structures within individual minds, and nowhere else ((Tononi, In preparation).

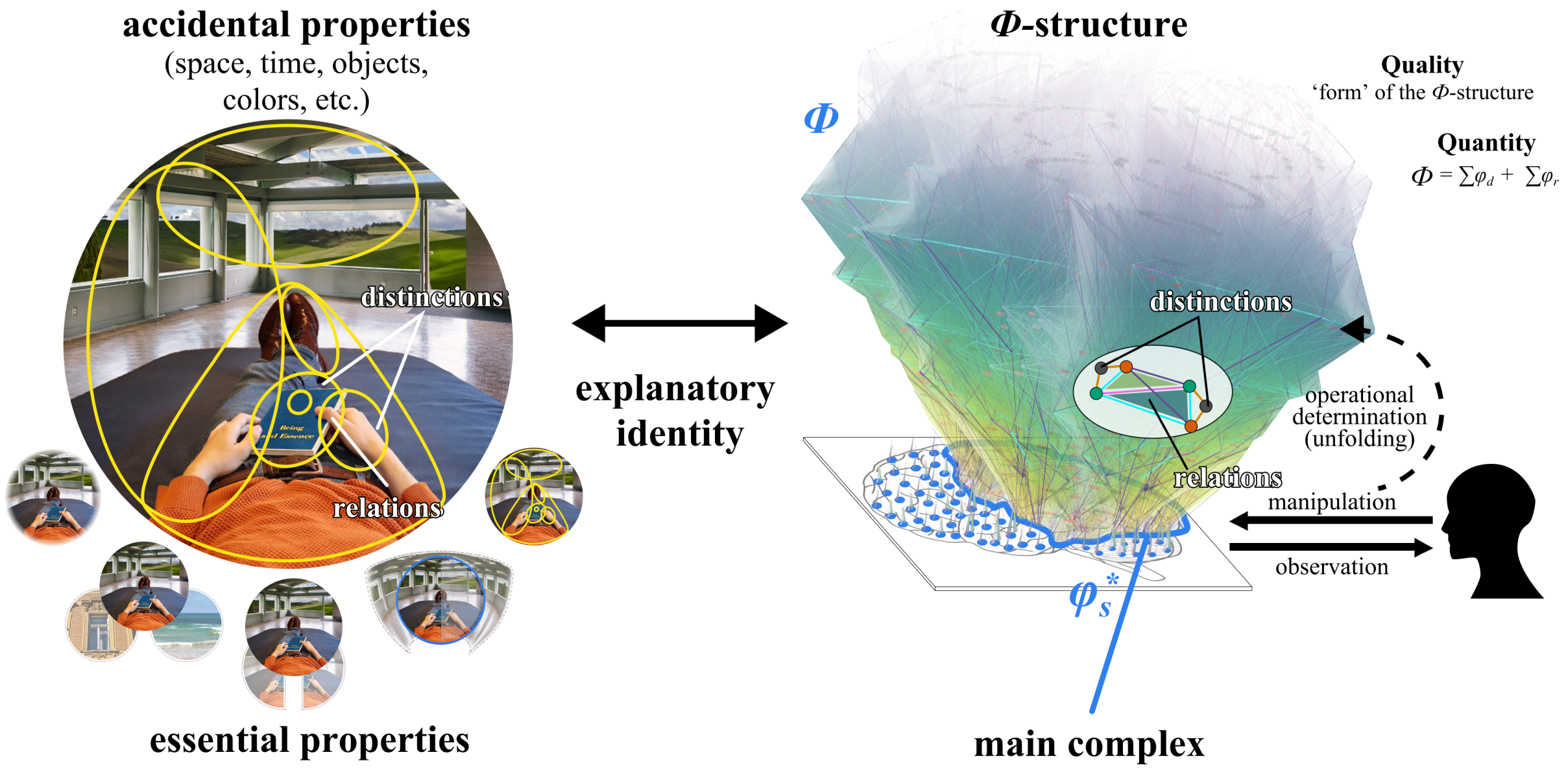

**Fig. 3** The explanatory identity of IIT. The central claim of IIT is that an experience (left side) is identical to a Φ-structure unfolded from a complex (right side). A Φ-structure captures the full cause–effect power of a substrate in a state unfolded according to the postulates of IIT. The quality of consciousness—the contents of an experience—is accounted for by sub-structures (Φ-folds) within the Φ-structure. Likewise, the quantity of consciousness is accounted for by the summed irreducibility (Φ value) of the components of the Φ-structure (distinctions and relations). The neural substrate of consciousness (brain, bottom right) is not viewed as "generating" the Φ-structure, but as the operational basis for unfolding it through observations and manipulations, say, by a researcher.

by converging and diverging connections, an arrangement that is well suited to high values of *Φ*. And indeed, both neurological and neurophysiological evidence, obtained through lesion, stimulation, and recording studies, indicates that posterior-central cortex likely constitutes the bulk of the neural substrate of consciousness (Boly et al., 2017; Siclari et al., 2017). By contrast, brain regions such as the cerebellum, which has many more neurons than the cortex and is indirectly connected to it but is characterized by a modular, feed-forward anatomy, can be lost due to tumor, accident, or surgery without affecting consciousness directly (Koch et al., 2016). IIT also predicts that the neural substrate of consciousness should cease to satisfy these properties when consciousness vanishes, as is the case in dreamless sleep or under anesthesia. This prediction has been confirmed by clinical studies showing that the efficacy of causal interactions within the cortex breaks down when consciousness is lost (Massimini et al., 2005; Pigorini et al., 2015).

Regarding the quality or content of consciousness, the explanatory identity of IIT predicts that the *Φ*-structure unfolded from the neural substrate of consciousness should account for the accidental properties of specific experiences. *Extensions*—cause–effect sub-structures unfolded from neural units connected in a grid-like manner, such as topographically organized areas of posterior cortex—can in principle account for the phenomenal properties that make spatial experiences feel extended (Haun & Tononi, 2019). *Flows*—cause–effect sub-structures unfolded from directed grids (as opposed to undirected ones)—should account for the phenomenal properties that make time feel flowing (Comolatti et al., 2025). *Conceptual hierarchies*—cause–effect sub-structures unfolded from pyramids of grids in posterior cortex—should account for the experience of objects, and so on for all other modalities of experience.

## 3.5 The principles of maximal and minimal existence

The information and exclusion postulates select the cause-effect state and the substrate of consciousness by invoking the *principle of maximal existence*, which plays a central explanatory role in IIT. In short, the principle says that, with respect to a requirement for existence, *what actually exists is what exists the most*. Thus, with respect to exclusion, the principle provides a sufficient reason for why an entity has the border it has rather than different borders (which would make it a different entity). For IIT, existence itself should be the reason: among a set of entities "competing" for existence over the same substrate, the entity that actually exists should be the one that lays the greatest claim to existence. Because what defines the quantity of integrated existence is irreducibility ($\varphi_s$), the entity that actually exists is the one that has the highest value of $\varphi_s$ ($\varphi_s$*). By the same token, candidate entities overlapping over the same substrate but with lower $\varphi_s$ are excluded from existence.[xi]

[xi] The same principle of maximal existence is applied to distinctions and relations, as well as to the grain of units, updates, and states, excluding from existence alternative entities over the same substrate. Because the irreducibility $\varphi$ is based on a measure intrinsic difference, it is sensitive to a tension between expansion and dilution ((Barbosa et al., 2020)). For example, given certain

In principle, to find out which entities exist intrinsically through observations and manipulations, we would have to measure the irreducibility of all candidate units (grains, updates, and states) and all candidate entities, for all possible partitions—over a universal substrate. In practice, assessing maxima of $\varphi_s$ exhaustively is out of the question, and one must resort to various approximations and simplifying assumptions. However, what exists maximally neither can nor needs to perform such exhaustive measurements—it does not need to "throw its weight around" to find out whether it should exist—whether it is indeed a maximum of irreducible, specific, intrinsic cause–effect power, and what its $\Phi$-structure might be. By the principle of maximal existence, only maxima of existence actually exist.

The principle of maximal existence (among candidate intrinsic entities, the one that actually exists is the one that would exist the most), combined with the principle of sufficient reason (there must be a reason why an intrinsic entity should be the one it is and not another) and with the principle of identity of indiscernibles (if two candidate entities have all the same properties, then they are one and the same intrinsic entity), imply the *principle of no free (intrinsic) existence*: a substrate unit can only support the cause-effect power of a single intrinsic entity. For example, assume that the neural substrate of consciousness—the "main" complex in our brain, turns out to correspond to a set of neuronal groups encompassing primarily posterior-central cortical areas (blue outline in **fig. 3**). Then, what exists here and now in physical terms is the $\Phi$-structure specified by this main complex in its current state (the large $\Phi$-structure depicted in **fig. 4**). And if the full cause–effect power of this main complex in posterior-central cortex corresponds to the large intrinsic entity that is my current experience, no further entities can co-exist over the same substrate: there cannot *also* exist some other entity corresponding, say, to one half of posterior cortex, or to one particular area within it, or to the entire cerebral cortex, or, say, to the entire brain. In other words, the full cause–effect power of the substrate of consciousness cannot co-exist with those of its subsets, supersets, and parasets[xii] (exclusion for sets). Otherwise, the full cause–effect power of a substrate would be multiplied "for free." Instead, what can also exist are other intrinsic entities, likely small ones, whose substrates do not overlap with the main complex. These mini-complexes with the small $\Phi$-structures they support are mere "ontological dust" (**fig. 4**)—unfolded from, say, groups of neurons arranged in partially segregated loops in prefrontal areas or in the cerebellum. These principles also imply that a substrate's intrinsic unit can only support the intrinsic cause-effect power parameters characterizing a substrate (including the strength of the connections among of the units and their organization, the internal mechanism of the units, and the amount of indeterminism), the size of the entity of maximum $\varphi_s$ will typically be intermediate between individual units and the entire substrate.

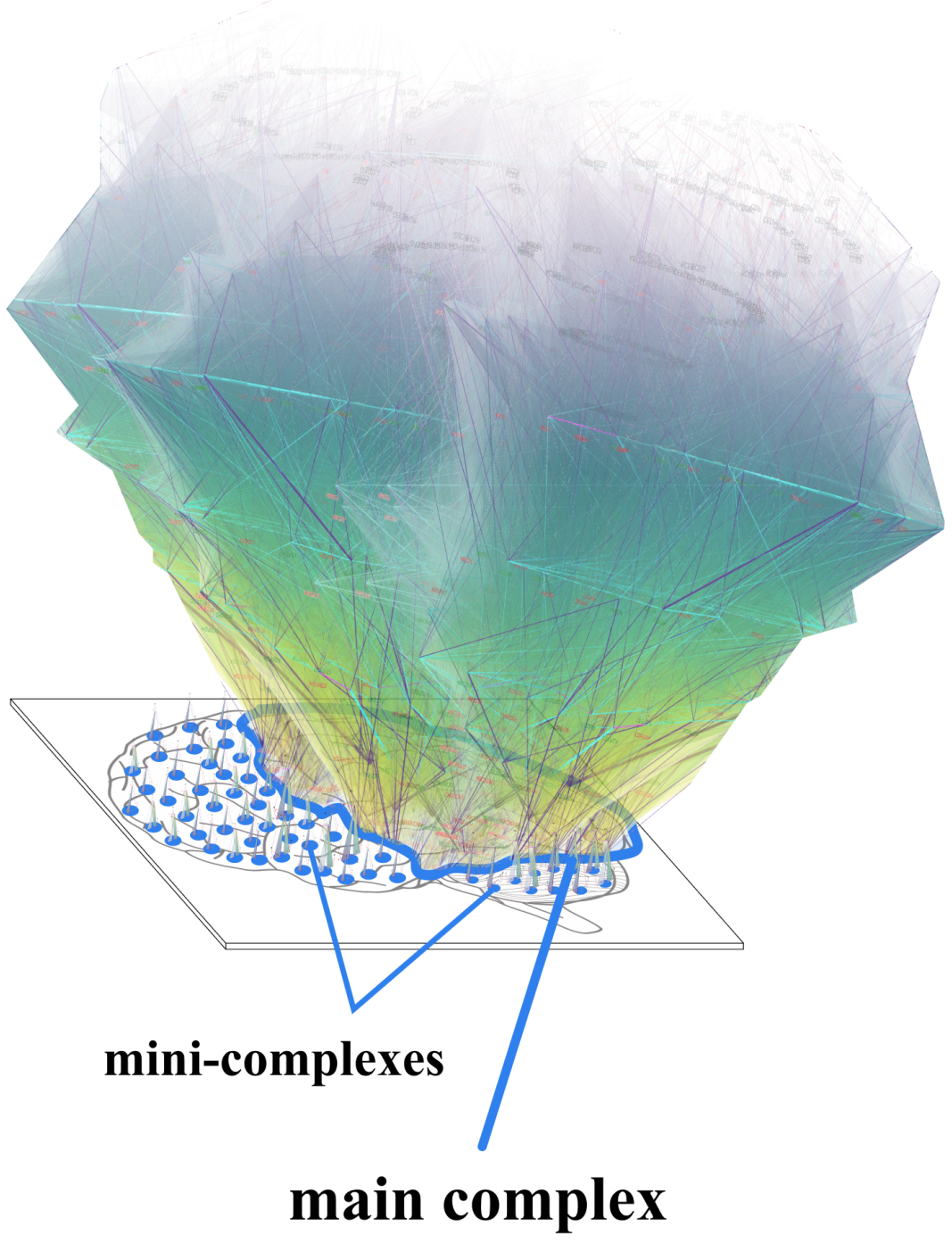

**Fig. 4** A neural substrate unfolded into one large $\Phi$-structure and many small ones. IIT predicts that the substrate of consciousness in the human brain, including primarily regions of posterior-central cortex, unfolds into an immense $\Phi$-structure, corresponding to an intrinsic entity (one's current experience), which excludes from intrinsic existence any entity unfolded from subsets, supersets, or parasets of that "main" complex. Subsets of units outside the main complex, say, in the cerebellum, subcortical nuclei, and regions of prefrontal cortex, specify mini-complexes that unfold into $\Phi$-structures in their own right, albeit minuscule ones—mere "dust" compared to the main $\Phi$-structure.

[xii] These are sets that include some units within the substrate and some units outside of it.

at its grain: it cannot *also* contribute to finer or coarse grains. For example, assume that the intrinsic units that constitute the main complex, and whose cause–effect power is unfolded into the entity's $\Phi$-structure, correspond to neuronal minicolumns (typically groups of 100 to 1000 neurons). Then no finer units (organelles, proteins, atoms, single neurons) or coarser units (hypercolumns, voxels, regions) can co-exist as intrinsic units over the same substrate, because their cause–effect power would be multiplied for free (exclusion for grains).[xiii] The maximal existence principle and the no free existence principle in IIT's ontology are somewhat analogous to the least action principle and conservation principles in physics, respectively (conservation principles are consequences of least action given some symmetries).

The principle of maximal existence is complemented by the *principle of minimal existence*. The principle states that, with respect to an essential requirement for existence, *nothing exists more than the least it exists*. The principle is offered by IIT as a good explanation for why, given that a system can only exist as one system if it is irreducible, its degree of irreducibility should be assessed over the partition across which it is least irreducible (the minimum partition). Moreover, a set of units can only exist as a system if it specifies both an irreducible cause and an irreducible effect, so its degree of irreducibility should be the minimum between the irreducibility on the cause side and on the effect side (as shown in **fig. 2**). Similarly, a distinction within a system can only exist as one distinction to the extent that it is irreducible, its degree of irreducibility should be assessed over the partition across which it is least irreducible, and it should correspond to the minimum between the cause and the effect side.

### 3.6 Intrinsic entities, extrinsic entities, and the great divide of being

We have argued that only a maximally irreducible substrate that unfolds into a $\Phi$-structure—an intrinsic entity—can account for the essential properties of phenomenal existence in physical terms. What must be emphasized now is that only an intrinsic entity can be said to exist intrinsically—to exist for itself, in an absolute sense. By contrast, if something has cause–effect power but does not qualify as an intrinsic entity, it can only be said to exist extrinsically—to exist for an external observer, in a relative sense (**table I**). And intrinsic, absolute existence is the only existence worth having.

To see why, imagine a world without any large intrinsic entity—a world in which every conscious being had fallen into dreamless sleep, or a world in which all conscious beings had died off, or the "early" universe as currently conceived, prior to the existence of stable planets that permitted life to evolve. In such a world of ontological dust, where nothing would feel like anything much, nothing would exist for which it would be true that something exists. In such a world, what would *existence* even mean? Schrödinger put it well: such a world would be "a play before empty benches, not existing for anybody, thus quite properly speaking not existing." The existence of something would be an inference nobody could draw, rather than an immediate and indubitable truth somebody would experience. Between intrinsic and extrinsic existence, then, passes the most fundamental of divides—the *great divide of being*. This is the divide between what exists in an absolute sense, in and of itself—namely conscious, intrinsic entities—and what only exist in a relative sense, for something else.

Substrates that do not unfold into intrinsic entities are merely "stuff"; they are substrates that can be observed and manipulated but that do not support maxima of cause–effect power and thus only have extrinsic existence. Stuff can range from assortments of macro or micro units that do not "hang together" (a random sample of air molecules), to aggregates that hang together loosely (an avalanche) or tightly (a rock).[xii] Indeed, not all the stuff that fails to qualify as an intrinsic entity is created equal. "Things"—stuff that hangs together tightly—can be considered *extrinsic entities*. These are substrates whose causal powers have high values of $\varphi_s$—higher than most of their subsets, supersets, and parasets—but they are relative maxima of cause–effect power rather than absolute maxima.

For example, my body hangs together well: it is a paradigmatic case of an "organism" that works as a unitary whole rather than as a set of isolated organs. It should thus have higher $\varphi_s$ than most of its subsets (say, my body minus some random assortment of organs), supersets (my body plus some garments), and parasets (my ears plus my shoes). However, if IIT is right, my body must be a relative maximum of $\varphi_s$ and inside my skull there must be a much higher, absolute maximum of $\varphi_s$ that is specified by the substrate of my current experience, presumably located within posterior-central cortical regions of my brain. This absolute maximum is an intrinsic entity and thus excludes my body from the realm of entities that exist absolutely, for themselves. In other words, since my body is a superset of my main complex, it is excluded from it—relegated to the realm of entities that only exist relatively, for an observer who can verify that they indeed hang together.

Bodies and organs, tables and rocks, stars and planets, and many other things that seem to hang together well are likely to unfold into extrinsic entities. Many objects of study in the special sciences, such as molecules, organelles, cells, organs, organ systems, organisms, groups, societies, and the like, which have proven to be good ways of carving nature at its joints, are also likely to correspond to relative maxima of $\varphi_s$. No matter how well extrinsic entities hang together, they do not exist for themselves. They only exist vicariously, from the perspective of some intrinsic entity, and so they do not properly exist. If I fall into dreamless sleep, my body will be breathing, my heart pumping, but in an absolute sense, there will be nobody there—in fact, not even a body—just an aggregate of much

[xiii] It also follows that, for the same complex, the constituents of an intrinsic unit cannot mediate the cause-effect power of other intrinsic units, and a background unit can only mediate the intrinsic cause-effect power of a single intrinsic unit.

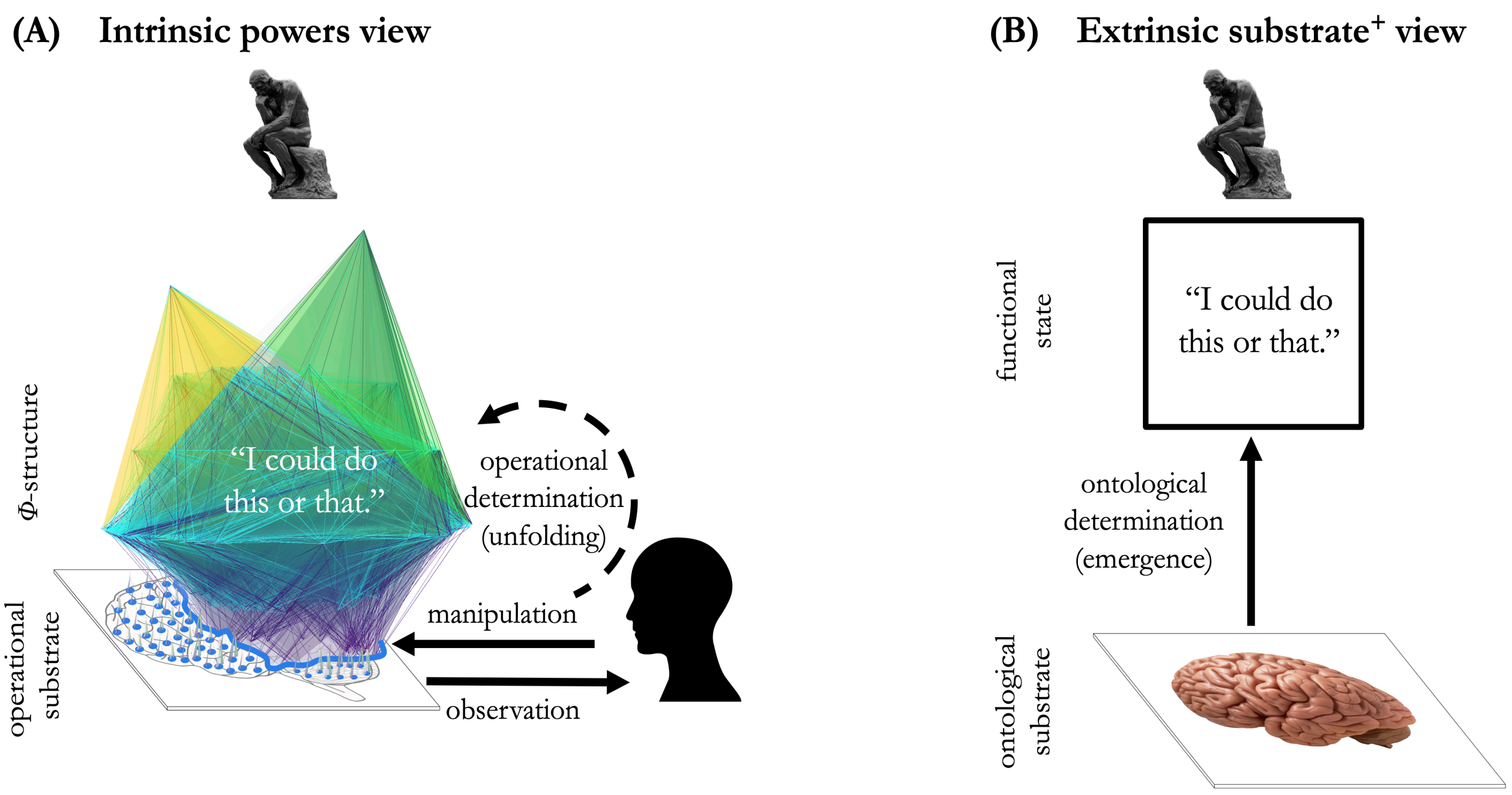

**Fig. 5** IIT's intrinsic powers ontology vs. extrinsic substrate$^+$ ontologies. Left side: In IIT, what exists intrinsically is not a substrate *as such* (e.g., a brain) but rather the fully unfolded intrinsic causal powers of a maximal substrate, captured by a $\Phi$-structure (top left). The substrate as such only exists extrinsically—as an operational basis that one can observe and manipulate (bottom left, faded gray brain). A $\Phi$-structure can be "vertically" *determined from* the substrate (e.g., by a researcher), but not *determined by* it; the maximal substrate unfolded into a $\Phi$-structure is what exists absolutely because it exists intrinsically. Right side: The ontology of IIT stands in contrast with various views, here bundled together as extrinsic substrate$^+$ ontologies. These typically view the microphysical substrate as ontologically fundamental (bottom right, colored brain), governing the vertical "emergence" of higher-level properties and constructs (top right). Functional states are thus ultimately *determined by* the microphysical substrate. Such states can be described as "real" inasmuch as they are necessary for explanation and are not reducible to the substrate owing to "multiple realizability," yet the microphysical substrate as such remains the ontological bedrock.

smaller entities.[xiv,xv] Yet for my conscious friend observing it, my body will continue to exist just as much as it did before I fell asleep.

### 3.7 IIT's intrinsic powers ontology versus extrinsic substrate$^+$ ontologies

As we have seen, IIT starts from phenomenal existence and defines physical existence operationally in terms of cause–effect power "all the way down," with no intrinsic residue, such as mass and charge. Accordingly, a substrate should not be thought of as an ontological or "substantial" basis—an ontological substrate—constituted of elementary particles that would exist as such, endowed with intrinsic properties. Said otherwise, existence is not constitution (here in the sense of the physical makeup of something, **table I**).

Instead, a substrate should be thought of as an operational basis—an operational substrate—from which we can unfold causal powers to reveal what actually exists. And what actually exists—according to the postulates of physical existence, based in turn on the axioms of phenomenal existence—are cause-effect structures unfolded from non-overlapping, maximally irreducible substrates, at a grain that maximizes irreducibility. Each complex is an intrinsic entity, fully characterized by its $\Phi$-structure, that exists absolutely because it exists for itself—as a conscious being (**fig. 5A**). In short, *the actual is the potential*, in the sense that what something is (the actual) is given by its powers (the potential).

We call this view IIT's *intrinsic powers ontology*. It differs from the assumptions usually taken for granted in neuroscience and biology, which align with what we will call *extrinsic substrate$^+$ ontologies* (**fig. 5B**). These ontologies are extrinsic because any convenient set of units, at any convenient grain, are considered to exist (although the "microphysical" level is typically considered more fundamental): brains exist, and within them different brain regions; neurons exist, and within them organelles and proteins; atoms exist, themselves constituted of elementary particles. Moreover, these extrinsic ontologies assume an ontological substrate because substrates are thought to exist as such, as an ontological basis endowed with fundamental physical properties, such as mass and charge, and following the "laws of physics."

A *substrate only* view would argue that all that exists is in fact the physical substrate (the Lucretian universe of "atoms and the void"), dismissing consciousness. *Substrate$^+$* or *emergentist*

[xiv] Collections of things that do not hang together extrinsically may still make sense from the point of view of a conscious observer (say, the set of unread books in Scandinavia), in whose mind they hang together as a highly interrelated content of experience.

[xv] Presumably, none of these small intrinsic entities will be conscious in a way that feels like much at all.

views hold that, *in addition* to the physical substrate, some kind of existence should be granted to properties and constructs, especially functional ones, that may "emerge" upon that substrate (hence the "plus").[xvi] These emergent properties and constructs are considered "real" for two main reasons. First, they are necessary for us to understand what organisms do (and, more generally, to understand any complex system). Second, the same emergent properties may be realized by multiple microphysical configurations, which makes these high-level properties not reducible to their specific substrate.[xvii] Thus, functions such as imagination, reasoning, and decision-making are granted existence because our behavior can only be explained by resorting to such constructs and because they may be realized differently in different brains (or even in computers). In substrate+ views, therefore, microphysical substrates exist *as such*, but high-level properties *also* exist in their own right, albeit in a less fundamental sense because they require the substrate to emerge.[xviii,xix]

The contrast with IIT's intrinsic powers ontology is fundamental. By IIT, when I am conscious, what actually exists is a complex unfolded into a large $\Phi$-structure, corresponding to my experience, and it exists at its particular grain. No subsets, supersets, or parasets of that complex *also* exist, just as no other grains *also* exist. Moreover, what actually exists is only the complex unfolded into a $\Phi$-structure corresponding to my experience, not *also* a substrate as such.[xx] Crucially, any content of my experience, including alternatives, reasons, and decisions, corresponds to a sub-structure within my $\Phi$-structure, not to a functional property emerging from my substrate.

This has an important consequence for free will: because I actually exist—as a "large" intrinsic entity—the neurons of my complex cannot *also* exist. They cannot exist as constituents of my complex, because what actually exists is not a substrate as such but the substrate unfolded into a $\Phi$-structure expressing its causal powers. And they cannot exist as small intrinsic entities in their own right because, if they specify a large intrinsic entity, they cannot also specify a set of smaller entities. Moreover, because my alternatives, reasons, and decisions exist within my experience—as sub-structures within an intrinsic entity—the neuronal substrates of alternatives, reasons, and decisions cannot also exist. If this picture is correct, it leaves no room for emergence or dualisms of any sort.

### 3.8 Existence and "vertical" determination

Despite their fundamental differences with respect to what actually exists, both IIT's intrinsic powers ontology and extrinsic substrate+ ontologies rely on what will be called vertical determination, albeit in two very different senses—operational and ontological.[xxi] A $\Phi$-structure can be fully and uniquely *determined from* a substrate of units in a specific state in an *operational* sense—through a set of observations, manipulations, and calculations (say, by a researcher). In this operational sense, the $\Phi$-structure can be unfolded "vertically" from the substrate's TPM based exclusively on the postulates of physical existence—no further ingredients needed (**fig. 5A**). What actually exists is the substrate unfolded into a $\Phi$-structure, because the substrate does not exist as such.

In extrinsic substrate+ ontologies, all emergent properties and constructs are fully and uniquely *determined by* the microphysical substrate in an *ontological* sense (**fig. 5B**). The emergence of high-level properties can be accounted for vertically by the properties of the substrate—no new physical properties or laws needed. Emergent properties and constructs exist—they are "real"—because they are not reducible to their substrate due to their explanatory role and their multiple realizability. The substrate actually exists as such; in fact, it must exist for non reducible high-level properties to emerge—to *also* acquire some kind of existence.

## 4 Causation: characterizing what causes what

Up to now, we have addressed the question "what exists?" We have aimed to account for our immediate, irrefutable phenomenal existence in terms of the cause–effect powers of a substrate in a state—say, a set of neuronal groups, some active and some not. We have shown that it is possible to vertically unfold a $\Phi$-structure that corresponds to a given experience.

---

[xvi] Here, "emergence" is used in the "weak" sense ((Bedau, 1997), which is commonplace in neuroscience and biology. Emergence is considered weak if high-level properties can ultimately be deduced (or demonstrated through simulations) from low-level properties, without requiring any new principles or "forces."

[xvii] *Not reducible* here means that a high-level function cannot be identical to its low-level, microphysical substrate because it is multiply realizable. The term *irreducible* is IIT is used instead to indicate that a substrate's cause–effect power is not the same as that of its parts.

[xviii] In the philosophical literature, such views are often grouped under the label of property dualism (as opposed to substance dualism) or non-reductive physicalism (a kind of closet dualism).

[xix] There is a variety of opinions about the status of phenomenal properties and their association with functional properties, including the idea that they may be "illusory."

[xx] We can say that, while an intrinsic entity *supervenes* operationally on its substrate (it can be vertically determined by observing and manipulating it, see below), ontologically is *supersedes* it (it is the entity that exists in place of the substrate as such).

[xxi] The notion of vertical determination here is related to *supervenience:* an experience can be said to supervene on a microphysical substrate (its supervenience basis), in the sense that any difference in the former is necessarily reflected in a difference in the latter. Either way, a substrate or supervenience basis are not assumed to exist as such but only as an operational basis from which what exists intrinsically can be unfolded. It should also be clear that vertical determination or supervenience have nothing to do with causation: a substrate should not be thought of as literally generating or "giving rise to" the cause–effect structure unfolded from it (such terms, just like the term "emergence" [see below], may be convenient in usage, but at the risk of being misleading).

We now turn to analyzing *actual causation*, namely, "what caused what?" (Albantakis et al., 2019; Grasso et al., 2021). This analysis can be formalized in a way that parallels the ontological analysis above, employing IIT's postulates to provide a self-consistent, quantitative account of many aspects of causation (Albantakis et al., 2019). We can distinguish the analysis of causal accounts, of causal processes, and of causal plasticity.

### 4.1 Causal accounts

To obtain a causal account, instead of considering the causal powers of a substrate in a state, we consider two successive occurrences—say, a set of unis in an earlier activity pattern and a set of units in a subsequent activity pattern. And instead of vertically unfolding a $\Phi$-structure, the analysis "*horizontally" unrolls* an $\mathcal{A}$-structure, which accounts for what caused what. In sum, while $\Phi$-structures capture existence in terms of *potential* causes and effects, $\mathcal{A}$-structures capture causation in terms of *actual* causes and effects.

As described above, to identify complexes and unfold $\Phi$-structures, we employ the criterion for *physical existence* ($0^{\text{th}}$ postulate) plus the five postulates of IIT. Similarly, to identify causal complexes and unroll $\mathcal{A}$-structures, we employ the criterion of *realization* plus the five postulates to characterize the transition between two substrate states (occurrences). Briefly, realization means that to establish actual causation, there must have been a transition between two occurrences, and the first occurrence must imply an increased probability of the second one. Realization requires that we have a repertoire of *counterfactuals*, corresponding to occurrences that could have happened but did not. Moreover, actual causation may look different from the two directions—not every cause corresponds to an effect, and not every effect corresponds to a cause. For example, the actual cause of a window being shattered may turn out to be a large rock, but this does not mean that a small rock hitting the window at the same moment did not have an effect.

The postulates of IIT appear in the analysis of actual causes and effects in the following form:

**Intrinsicality** means that actual causes and effects must be evaluated from the intrinsic perspective of the occurrence—relative to it (e.g., what cause makes the most difference to it).

**Information** means that cause and effect states must be specific—in this case corresponding to the states that actually occurred.

**Integration** means that a cause or effect must be irreducible—otherwise there would be two or more independent causes or effects rather than one. This irreducibility is measured by causal strength $\alpha$, which is formally analogous to integrated information $\varphi$.

**Exclusion** means that cause and effect must be definite—every occurrence must have one antecedent cause and one subsequent effect. The cause and effect are chosen based on maximal irreducibility ($\alpha$*).[xxii]

**Composition** means that the cause and effect must be structured—we should consider sub-causes, sub-effects, and how they overlap. Accordingly, once the overall cause and effect involved in the transition between two occurrences have been identified, its *$\mathcal{A}$-structure* can be unfolded by assessing causes and effects for each of its subset and how these overlap over units. This yields an optimal or 'main' *causal account* of a single transition.[xxiii]

### 4.2 Causal processes

An analysis of *causal processes* further characterizes the horizontal unrolling of causal accounts across many transitions over different substrates by evaluating their congruence (how an effect of a previous transition overlaps with the cause of the next transition).[xxiv] A causal process typically has borders, a beginning, an end, and an internal structure, and it can occur at a macro rather than at a micro scale.

For both causal accounts and causal processes, maxima can be considered at multiple levels. For example, in some cases the strict maximum of irreducibility over a substrate may occur at a "meso" grain and across multiple "narrow" accounts and processes, but near-maximal "macro" grains and "broad" accounts and processes may reveal relevant levels of causal organization, similar to how extrinsic entities reveal relevant levels of extrinsic existence. Such near-maximal, broad causal processes may be said to "stream together" well, similar to how extrinsic entities "hang together" well. It is also important to consider how causal accounts and causal processes relate to intrinsic entities—whether they originate within or without.[xxv]

### 4.3 Causal plasticity

Finally, an analysis of *causal plasticity* is needed to determine how causal processes change the transition probability matrix (TPM) that characterizes a substrate. IIT's *principle of becoming* (powers become what powers do) (Tononi, In preparation) assumes that the TPM changes based on nothing else than the sequence of states that actually occur.[xxvi]

---

[xxii] Like integrated information, actual causation is based on probabilities and is sensitive to the tension between expansion and dilution (see footnote[x]). For example, a set of units may have a cause over a large number of units (the total cause), but only a few of these may turn out to be the actual cause.

[xxiii] The key difference in computing causal strength $\alpha$ and integrated information $\varphi$, and from there $\mathcal{A}$ and structured integrated information $\Phi$, is that $\alpha$ picks *actual* cause and effect states—those that actually occurred—whereas $\varphi$ picks *potential* cause and effect states—those that maximize intrinsic existence.

[xxiv] This work is currently in development.

[xxv] A causal process can also be interpreted as a realization, over many transitions, of the causal potential of an entity in a state, given some background conditions.

[xxvi] Briefly, powers are nothing but conditional probabilities reflected by the TPM, and as such, they too update when unit states update. The plasticity of powers ultimately provides the sufficient reason for the powers to have become what they have become, rather than being what they are arbitrarily (for no reason). The precise way in which powers are updated—how *actions beget*

Establishing what caused these changes can be accomplished by considering a cause set and an effect set of units, measuring the changes in the TPM of the effect set across sequential transitions, and weighing them by the causal strength *a* of the associated cause and effect. Through an appropriate selection of cause and effect set and of relevant sequences, such an analysis would reveal what caused the changes in the substrate, ultimately yielding a causal account of the substrate itself.

### 4.4 Causation, prediction, and "horizontal" determination

There is a history of skepticism towards causation in both science and philosophy. It has often been assumed that the business of science is to predict accurately, not to postulate unobservable causes. All we can observe, in the end, are just conditional probabilities across occurrences. However, when dealing with complex systems, such as biological ones, prediction is problematic because we cannot rely on simple "laws." Instead, to figure out how a complex system works, we need systematic interventions in addition to observations, ideally yielding an accurate "causal model." For example, neuroscientists have long recognized that "correlation is not causation" and that, to understand the brain, we ultimately need an accurate "causal model" of neurons in a network. With such a model, we can predict the next state of individual units from the state of their inputs. If we do so for every unit, no extra work is needed to predict the state of the entire network. In other words, first-order prediction is all that is needed for high-order prediction (Grasso et al., 2021).

Even so, it is critical to understand that *prediction is not the same as causati*on (**table I**). Consider again a network of neurons. Prediction can be understood as *horizontal determination*—the ability to unroll the future state of a system's units based on knowledge of their past state and of the units' mechanisms. Instead, causation should be understood as the ability of a mechanism to take or make a difference, as demonstrated through observation and manipulation.

On this basis, it is easy to show that the two notions can be dissociated. For example, it is true that if something is caused according to a causal model, then it can be predicted, but the converse is not true: something can be predicted but not caused. Consider two identical, parallel chains of domino stones. Triggered by a common event, dominos in both chains start to topple. The fall of one domino in one chain reliably *predicts* the subsequent fall of the next dominos in both chains, yet it only *causes* the fall of the domino in its own chain (examples can also be drawn from neuroscience) (Grasso et al., 2021). Moreover, high-order predictions follow from 1st-order predictions: if the next state of neuron A is predicted, and similarly that of neuron B, then necessarily the next state of AB is also predicted. Not so for high-order causation: there may be many 1st-order causes (such as A and B) and no high-order causes (AB), or there may be a high-order cause (AB) in the absence of 1st-order ones (A and B). A similar dissociation can be shown for macro causation vs. macro prediction. Knowing the state of all relevant micro-units (say, molecules) over their micro updates is sufficient to predict the state of macro units (say, neurons) over macro updates. However, causal analysis may show that actual causes and effects are maximal at the macro level rather than at the micro level (Grasso et al., 2021).

### 4.5 Causal closure and causal exclusion in extrinsic substrate+ views

In **section 3.7**, we discussed differences between IIT's intrinsic powers view and standard extrinsic substrate+ views with respect to what exists. How do these ontologies differ with respect to what causes what?

As we have seen, extrinsic substrate+ ontologies assume that the microphysical substrate exists as such, endowed with physical properties such as mass and charge. Because it exists as such, it also causes as such.[xxvii] Emergent properties are also assumed to exist, albeit in a less fundamental manner, because they depend on the existence of the microphysical substrate. As we have also seen, their existence is justified by the role they play in explanation and by multiple realizability. The same logic can be applied to what might be called "emergent causation." It seems that we cannot avoid resorting to high-level causes to understand our behavior. For example, "he caused a lot of hardship to a lot of people" can only be understood at the high level of agent causation. Moreover, high-level causation can be multiply realized. Could we not accept, then, that emergent causes are *also* causing, albeit in a less fundamental sense than microphysical causes?

Most think not. While extrinsic substrate+ views are not overly worried by ontological proliferation (emergent properties co-existing with their substrate), there is widespread worry about causal proliferation or "overdetermination." The intuitive reason is based on the notion of causal closure of the physical world: microphysical causation is sufficient for determining everything that happens (horizontal determination)—in other words, no additional causal influences need to be invoked to determine what happens next.[xxviii] High-level properties and constructs are vertically determined by their microphysical substrate, both before and after causation takes place. Therefore, there is nothing high-level properties and constructs can do causally above and beyond what has already been caused at the microphysical level. In short, a principle of causal exclusion is applied according to which microphysical causation excludes higher level causation (Kim, 1998).

---

*powers* (the plasticity function, coupled with the activity function—how *powers beget actions*)—must be consistent with the principles and assumptions of IIT and is discussed elsewhere ((Tononi, In preparation)

[xxvii] This holds inasmuch as causation is acknowledged and distinguished from prediction ((Grasso et al.)).

[xxviii] This holds unless one subscribes to interactionist dualism, according to which mind can have effects on matter ((Popper & Eccles)), or to strong emergence, according to which new physical forces or laws may come into being at higher levels of organization.

### 4.6 Only what exists can cause: intrinsic and extrinsic causes

IIT's intrinsic powers ontology, augmented by its causal analysis, leads to a fundamentally different view: *only what exists can cause*. More precisely, only what exists absolutely, which is to say intrinsic entities, can be properly said to cause. Conversely, extrinsic entities, which only exist relative to conscious beings, cannot properly cause anything because they do not exist absolutely. We may call a causal account or process originating from an intrinsic entity an *intrinsic cause* and one ending in an intrinsic entity an *intrinsic effect,* to be distinguished from extrinsic causes and effects (**table I**).

**Table I: Some key contrasts in IIT's ontology**

| | |
|---|---|
| constitution $\neq$ **existence** | **intrinsic entity** (an absolute maximum of $\varphi$) |
| | **extrinsic entity** (a relative maximum of $\varphi$) |
| prediction $\neq$ **causation** | **intrinsic causation** (begins or ends in an intrinsic entity) |
| | **extrinsic causation** (does not begin or end in an intrinsic entity) |

To exemplify, consider two occurrences in which my hand is raised. In one case, my hand is raised because I have decided to do so voluntarily—say, to ask a question. Causal process analysis would establish that the *distal* cause—where the causal process begins—resides in a large intrinsic entity, the large $\Phi$-structure corresponding to my experience of deciding that is supported by a neural substrate in posterior-central cortex. The causal process originating there activated motoneurons in my motor cortex, neurons in the spinal cord, and eventually muscles that raised the arm. The distal cause is therefore an intrinsic cause because it occurs within an intrinsic entity.[xxix] Because an intrinsic entity exists for itself, when it produces certain external effects, it can be considered an *autonomous agent*; and when it is sensitive to certain external causes, it can be considered an *autonomous perceiver*.

In a second case, my hand is raised involuntarily following some burst of activity in the motoneurons in my spinal cord while I am deeply asleep and unconscious. Causal process analysis would establish that the distal cause is spread out among small entities of neuronal "dust" in my spinal cord (no large complex with a large $\Phi$-structure but many small ones). In this case, the distal cause remains extrinsic because it did not occur within an intrinsic entity that would serve as an autonomous agent or perceiver.

At first, the intrinsic powers view may seem counterintuitive. For instance, with respect to existence, it implies that bodies and rocks only exist relative to a conscious observer and manipulator, as extrinsic entities, but not in an absolute sense, as intrinsic entities existing for themselves. The situation is similar with respect to causation. Take the case of a rock rolling down a hill and crashing into an empty car. Causal analysis would be able to pinpoint the rock as an extrinsic cause of the car being crashed, and the car being crashed was its effect (whether the casual accounts and processes are maximal or near-maximal). However, both the cause and the effect would be extrinsic, because neither the rock nor the car exist as intrinsic entities, but only as aggregates of much smaller intrinsic entities that hang together tightly. Properly speaking, it is a case of dust causing dust.

In a fundamental sense, then, because there is 'nothing it is like' to be a rock or a car—neither is an intrinsic entity—the rock cannot be an intrinsic cause and the crashed car cannot suffer an intrinsic effect.[xxx] This conclusion does not mean that an extrinsic causal analysis yielding the rock as an extrinsic cause and the crashed car as an extrinsic effect would be meaningless, or that one should cease speaking about the rock causing the crash. Just as relative maxima of $\varphi$s identify things that "hang together well," such as rocks and cars, bodies, and spinal cords, maxima of $\alpha$ can identify causal accounts and processes that connect extrinsic entities and "stream together well."

The requirements for high $\mathcal{A}$ are very similar to those for high $\Phi$ because they are similarly based on the postulates of IIT. Therefore, a distal cause that begins or an effect that ends in a $\Phi$-structure of high $\Phi$ will typically have high $\mathcal{A}$. In other words, the more an entity is conscious, the more strongly it can cause, and the more strongly it can bear an effect. Thus, raising my hand to ask a question not only has an intrinsic cause—my conscious decision to ask—but it will likely have a strong cause. Instead, a near-random coincidence of firing in the spinal cord, which just so happens to raise my hand when I am asleep, will have a comparatively weak cause, in line with the spinal cord not being conscious—a mere aggregate. In general, we should expect a proportionality between the $\Phi$ value of a $\Phi$-structure, and hence of $\Phi$-folds within it, and the $\mathcal{A}$ value of a causal account. As we will now argue, intrinsic causes and effects underlie free will and the ability to choose.

## 5 Requirements for free will

The main thesis of this essay is that we do have genuine free will and not just "free will" as a convenient manner of explaining our actions. Before substantiating this claim in the IIT framework, let us return to the simple scenario introduced at the beginning and examine some requirements for free will in more detail. It is generally recognized that free will requires

[xxix] More precisely, the distal cause must be specified by a subset of the same macro units that compose a $\Phi$-structure, over the same macro state.

[xxx] Each mini dust entity would affect some mini dust entity.

**Table II: Requirements for free will**

**Consciousness**

- **Alternatives**: I must be able to envision multiple courses of action. (*freedom of imagination*)
- **Reasons**: I must be able to choose based on reasons. (*freedom of evaluation*)
- **Decision**: I must be able to decide and intend an action. (*freedom of decision or will*)
- **Control**: I must be able to cause, control, and execute an action. (*freedom of execution*)

consciousness. But consciousness is not enough (**table II**). In other words, there cannot be free will without consciousness, but there can certainly be consciousness without free will. A proper free will scenario requires that I can envision alternative courses of action among which I can choose (freedom of imagination); that I can weigh reasons to choose among alternatives (freedom of evaluation); that I can decide and intend an action (freedom of decision or will); and that I can cause, control, and execute an action to carry through my decision (freedom of execution).

*Freedom of imagination*. Say I am driving on a long stretch of a deserted highway as if on autopilot, immersed in the music from the radio. While I am conscious and in control of the car, I do not envision alternative courses of action, so freedom does not come into play (until, of course, I wonder whether I should turn left to go home or right to go for a snack). The same applies if I were to do what I am told without ever considering that I might do otherwise. Note that it would not matter at all whether, due to some accumulation of indeterminism, two possible trajectories were open to me. What matters is only that I am conscious of a fork, so to speak, in my mental road—that I realize I have a choice.

*Freedom of evaluation*. Say I do see alternatives but choose automatically among them, almost reflex-like, because I have been thoroughly indoctrinated about what is the right choice. If I never considered the reasons why I should choose this rather than that, I am not free. I may consider my reasons quickly or go through a long, agonizing process, but as long as I have considered my reasons, I have freedom of evaluation.[xxxi] Usually, when a choice is difficult, I am also acutely aware that the choice is ultimately up to me and up to me only—that I am free to choose. In fact, awareness of the very concept of freedom adds to the poignancy of the choice I must make.

*Freedom of decision or will*. Say I am asked to vote for or against a resolution that is unpopular but ethically well grounded. Normally, I would consider my alternatives, go through my reasons, and decide to vote accordingly. In this case, I would have decided as I intended and exercised my free will. But now imagine that, just when I am about to decide, somebody intervenes and, by targeting an intense magnetic field to neurons involved in decision-making in my brain, causes them to fire and "decide." In such case, the decision cannot be considered an expression of my own free will. Similarly, even though I may have pondered alternatives and gone through my reasons, if in the end I succumb to a sudden fit of rage, my decision was not freely willed.[xxxii] Of course, if I am unable to decide because I am paralyzed by doubt, I also lack will.

*Freedom of execution*. Say I see alternatives (should I drink alcohol or not?), go through my reasons (for the sake of my health, I should not), and decide to quit (I will not drink). However, if my resolve is thwarted by an overwhelming push to transgress, I may fail to carry out my decision and cannot control my action. In that case, I lack freedom of execution. Note that this kind of "push" can be considered extrinsic, even though it may originate within my brain but outside the main complex, because I am not able to control it through my will. The same is true, of course, if I cannot carry out my decision because I am prevented by somebody or something external to me.

# 6 Genuine free will: the intrinsic powers view

We will now consider how we can account for the free will scenario introduced at the beginning of this essay in view of IIT's ontological foundations and its characterization of actual causation.

## 6.1 I have alternatives—not my neurons

As summarized earlier, by IIT, all contents of experience correspond to sub-structures within a cause–effect structure—to $\Phi$-folds within a $\Phi$-structure. This applies not only to the experience of space, time, and objects, but also to conscious thoughts and feelings of any kind. For example, the conscious

[xxxi] My choice can still be considered free even if I do not reconsider my reasons in any depth because I have done so in the past and molded my own character, beliefs, and values through self-changing actions (see below). However, I must be willing and able, in principle, to reconsider them in view of new circumstances.

[xxxii] We will not consider simple urges, say, extreme hunger or thirst, as adequately representative of a free will scenario, which involves the conscious consideration of alternatives and reasons. While an urge followed by a decision captures the "will" portion of free will as well as freedom of execution, it does not capture freedom of imagination and evaluation.

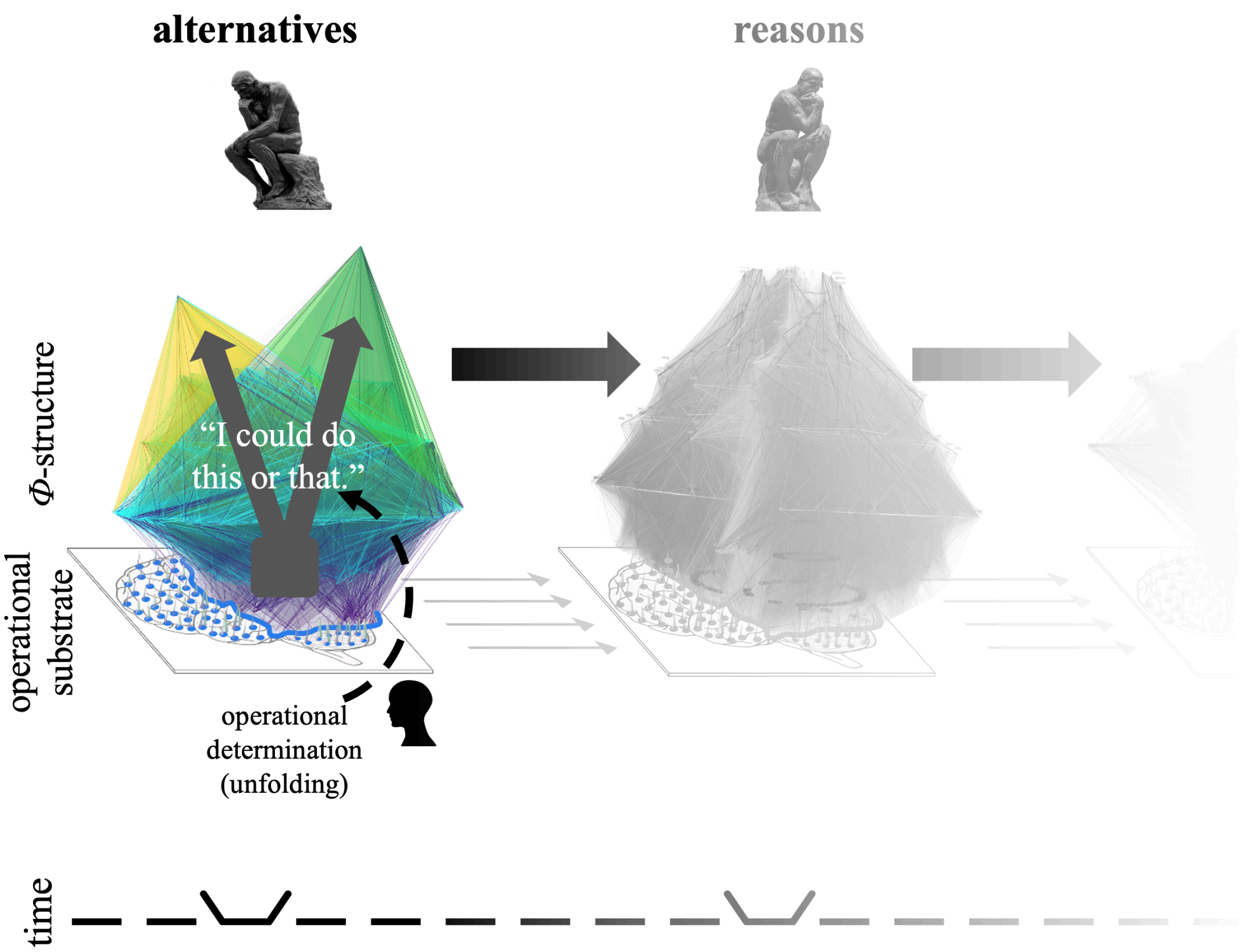

**Fig. 6** I have alternatives—not my substrate. A genuine free will scenario starts as a thought comprising two (or more) alternatives—a choice. This thought is a "fork in the mind," as depicted by the two prominent sub-structures (yellow and green) in the overall $\Phi$-structure corresponding to the experience.

content of a thought might correspond to the higher (more abstract) levels of a conceptual hierarchy, bound by relations to $\Phi$-folds corresponding to the sound of certain words. Conscious alternatives, too, are $\Phi$-folds within the $\Phi$-structure corresponding to an experience. Right now, I may conceive of two alternative courses of action as a silently verbalized thought. Such a thought would correspond to a large $\Phi$-fold that likely includes several conceptual hierarchies, extensions, flows, and high-level concepts such as the self. Different $\Phi$-folds may come into being when I go back and forth between possible courses of actions and their consequences, recall episodes from memory, and so on.

At some point, I will end up with a thought that binds two alternative possibilities into a *choice* that is available to me: I have arrived at a fork in my mind (**fig. 6**). That conscious choice exists phenomenally, and it must be accounted for, in physical terms, by a $\Phi$-fold within the $\Phi$-structure corresponding to my experience. In fact, as argued in **section 3.7**, that $\Phi$-structure unfolded from my main complex, and the $\Phi$-fold within it, is *all* that exists in an absolute sense—that is, for itself. The substrate and the neurons constituting it do not exist as such, but only relative to an observer—as an operational basis for unfolding causal powers. And, as argued in **section 3.7**, the neurons of the main complex do not exist unfolded either (as minuscule intrinsic entities). This is because, over a given substrate, what exists is only the entity that exists the most, which is the much larger intrinsic entity corresponding to my experience. So it is *I* who exist—not my neurons. And it is my thought of alternative actions—my conscious choice—that exists here and now as a $\Phi$-fold in the $\Phi$-structure of my experience—*a fork in my mind* (a "Y"). What exists is not a set of alternative trajectories that my brain may follow—"a fork in the road" (a "-<"), say, due to indeterminism (see **sec. 8**).

### 6.2 I have reasons, decisions, and control—not my neurons

The same logic applies to subsequent moments of the free will scenario (**fig. 7**). A reason—a thought about what might justify a choice—exists as a $\Phi$-fold within a current $\Phi$-structure. Reasons likely correspond to very complex $\Phi$-folds, and so do conscious beliefs, values, and the feeling that my own self is involved in a decision. According to IIT, all this content comes into existence when it is thought consciously, and it vanishes the moment it ceases to contribute to my current experience. Similarly, the feeling of making and then having made a decision, the feeling of my intention to carry it out, and finally the feeling of control—whenever they come into being as contents of my current experience—exist as large $\Phi$-folds within the corresponding $\Phi$-structure. And again, whenever I experience such contents, it is *I* who exist and *I* who have reasons, decisions, intentions, and control—not my neurons, which do not *also* exist. The feeling of *agency*—that I was the one responsible for executing an action—is also typically associated with free will, though under certain circumstances one may incur illusions of agency.

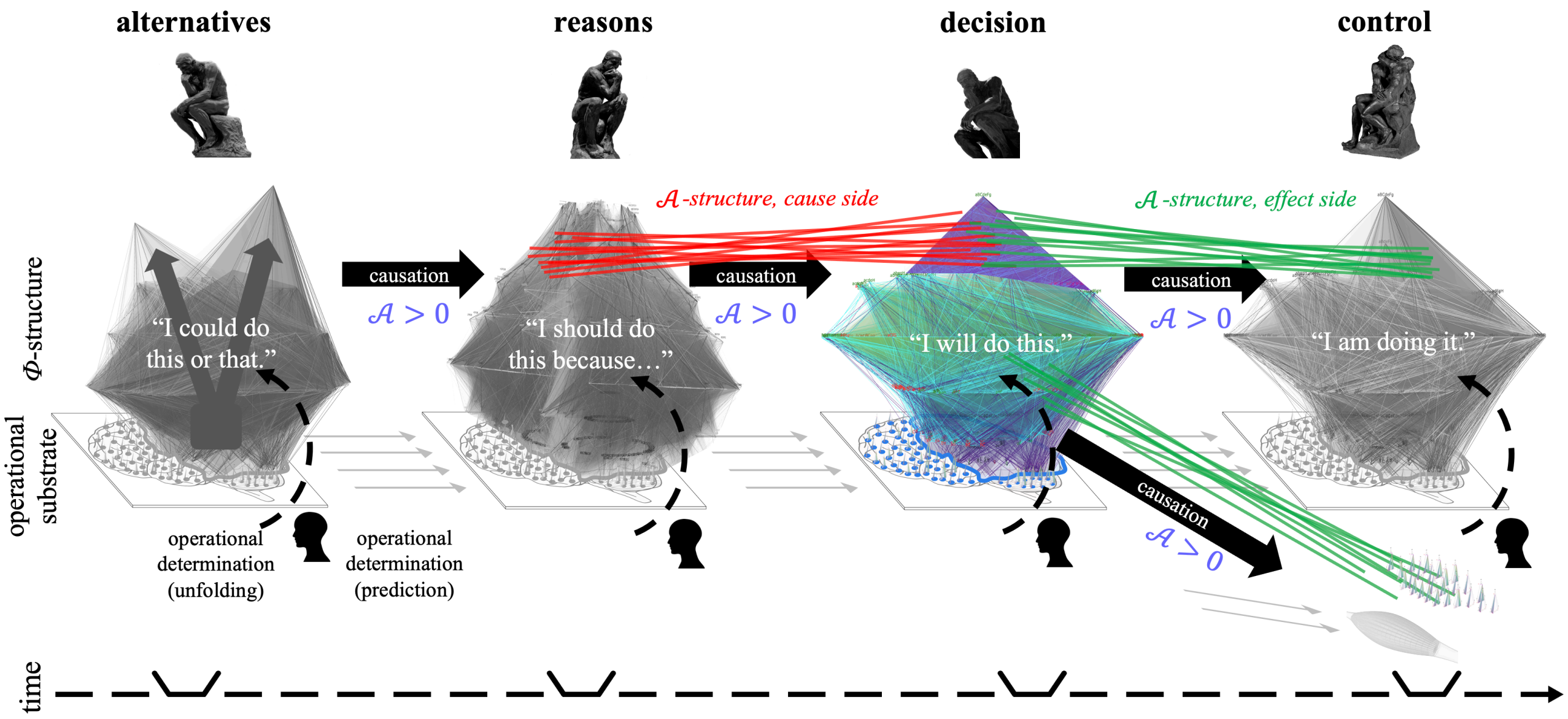

**Fig. 7** The intrinsic powers view of IIT: only what exists can cause. My experience of alternatives, reasons, a decision, and the feeling of control correspond to $\Phi$-structures unfolded from my substrate. At any one moment (say, the decision moment), one such $\Phi$-structure (the one in color) is what exists intrinsically. And as such, they are what can genuinely cause. A causal account of the decision is depicted by the red cause lines and the green effect lines, representing the supported $\mathcal{A}$-structures. On the cause side, one sub-structure in the reasons $\Phi$-structure is the actual cause (with structure integrated causation $\mathcal{A} > 0$) of a sub-structure in the decision $\Phi$-structure. On the effect side, the actual effect (with $\mathcal{A} > 0$) is the movement enacted by my motor system (indicated by the muscle, bottom right), which I control, as well as the feeling that "I am doing it."

We can now consider what caused my decision and my subsequent action. Recall that the question "what caused what?" is addressed by "unrolling" an $\mathcal{A}$-structure, which causally links two occurrences (in this case, two consecutive neural firing patterns in my brain).

Let us focus on the conscious moment at which I decide that "I will do this" (**fig. 7**). The experience corresponds to a $\Phi$-structure unfolded from the main complex—say a set of neuronal units in posterior-central regions of my cerebral cortex, some active and some not. The decision corresponds to a $\Phi$-fold within the $\Phi$-structure specified by a subset of those units. The previous conscious moment, say, the experience of a decisive reason why I should choose one alternative and not the other, corresponded to a different $\Phi$-structure, unfolded from a different activity pattern over some other subset of main complex units.

In principle, the $\mathcal{A}$-structure that accounts for what caused my decision could now be calculated (**fig. 7**). We would have to assess the actual cause of the state of the main complex units specifying the $\Phi$-fold corresponding to my current decision. As illustrated schematically in the figure, this actual cause and associated causal account is likely to involve many of the neuronal groups that, a moment before, had specified the $\Phi$-fold corresponding to my experience of a decisive reason. Just as a $\Phi$-fold within $\Phi$-structures of high $\Phi$ is likely composed of many distinctions and relations, the causal account linking my current decision $\Phi$-fold to my previous reason $\Phi$-fold will likely be a large $\mathcal{A}$-structure composed of many sub-causes, sub-effects and their relations.[xxxiii] The causal account on the effect side of my decision would behave similarly. The neuronal units specifying my decision $\Phi$-fold will have strong actual effects over the next state of many units. Some of these will then trigger further effects downstream, eventually leading to the activation of motoneurons and muscles, ending in an action such as raising my hand. Again, the $\mathcal{A}$-structure on the effect side will typically be large because, by originating within a $\Phi$-structure of high $\Phi$, it will be composed of many sub-effects and their relations.

By unrolling "what actually caused what" in this way, we would thus conclude that, in physical terms, my decision $\Phi$-fold was caused primarily by a decisive reason $\Phi$-fold that allowed me to choose among several alternative $\Phi$-folds; in turn, my decision $\Phi$-fold caused my subsequent action. And because the causal account of my decision ends within an intrinsic entity on the cause side (what caused it) and begins from an intrinsic entity on the effect side (what it caused), my decision exists intrinsically, has an intrinsic cause, and produces an intrinsic effect. Recall the conclusion from **section 3.7** on ontology: I exist—not my neurons. And recall the conclusion from **section 4.6** on actual causation: only what exists can cause. So I exist and cause—not my neurons.

[xxxiii] The strength of actual ($\alpha$) and potential causes ($\varphi$) tend to be correlated. In a relatively isolated system (say, my brain when dreaming or thinking), the actual cause of a mechanism in a state will tend to line up with its intrinsic cause.

### 6.3 A computer simulating my brain (or my functions) would not genuinely exist nor cause

This case for genuine free will may be further illuminated through a contrast with digital computers. A digital computer may be programmed to be functionally equivalent to a system it is simulating, yet specify $\Phi$-structures that are radically different, generally trivial, and largely indifferent to what is being simulated (Findlay et al., 2024). For example, the behavior of a small system of four logic gates (named PQRS) can be simulated by a simple universal computer constituted of one hundred seventeen logic gates (named PQRS'). PQRS' is *functionally equivalent* to PQRS: in terms of what it does, PQRS' indeed does the same as the system PQRS it is simulating. However, PQRS is a single complex with $\varphi_s>0$, unfolding into a $\Phi$-structure composed of numerous distinctions and relations. By contrast, the much larger substrate PQRS', the computer taken as a whole, has $\varphi_s=0$, being fully reducible. Subsets within it, such as memory registers and the clock, have $\varphi_s>0$. However, the trivial $\Phi$-structures unfolded from these subsets have nothing to do with the $\Phi$-structure of the system PQRS that is being simulated. In fact, those $\Phi$-structures would be the same no matter which system were being simulated by PQRS'. Thus, the system PQRS and the computer PQRS' simulating it are *functionally equivalent* but not *ontologically equivalent*: PQRS is a single intrinsic entity (albeit small), whereas PQRS' is just an aggregate of "dust."[xxxiv]

In short, the computer as a whole does not exist intrinsically—for itself—it is not an intrinsic entity. It only exists for us as a useful aggregate of ontological dust that can perform certain functions. But then, if the computer does not exist, it cannot cause. In fact, causal analysis would reveal many separate causal accounts linking, say, individual transistors constituting a memory register—another case of dust causing dust (**fig. 12** in the Supplementary Information).

# 7 The impossibility of genuine free will: the extrinsic substrate$^{+}$ view

Let us now contrast again IIT's intrinsic powers view with the extrinsic substrate$^{+}$ view. The latter comes naturally to most people, whether as naïve realists or as sophisticated scientific realists. In this view, physical substrates exist *as such*: my body certainly exists as a physical substrate, and so do my head, my brain, my cortex, its posterior-central regions, or any particular cortical area. The units we can observe and manipulate with the tools of science also exist *as such*: neurons certainly exist, and so do ion channels and even atoms, at any spatial and temporal scale we may be interested in. Furthermore, the substrate$^{+}$ view allows for "emergence": in addition to the substrate, high-level properties and constructs *also* exist, though they are vertically determined by the substrate. For some, this kind of emergence may offer a feeling of free will that is "good enough"—the "variety of free will worth wanting" (Dennett, 1984).

### 7.1 No free will? Ontological and causal micro determination

However, the extrinsic substrate$^{+}$ view is ultimately incompatible with genuine free will.[xxxv] To illustrate why, let us revisit once more the free will scenario in light of ontological and causal micro determination (**fig. 8**). My brain, constituted of neurons and their connections (and ultimately of elementary particles), is assumed to exist as such—as a fundamental physical substrate endowed with intrinsic properties such as mass and charge. That substrate in its current state ontologically determines all associated high-level emergent properties and constructs ("functional states" in the figure), which might correspond to, say, imagining alternative courses of action.

In this view, the next state of the neurons constituting my physical substrate is causally micro determined by their previous state, at least over the few seconds of the free will scenario. As illustrated in **fig. 8**, my neurons will switch from state to state in a way that is causally determined, in microphysical terms, by their inputs (their previous state) and their mechanisms. Inevitably, the same way my experience of imagining alternatives is vertically determined by a neural state, my experience of a decisive reason favoring the first alternative is vertically determined by a neural state, as is my experience of deciding for one alternative.

The extrinsic substrate$^{+}$ view emphasizes that the microphysical level is sufficient to predict what I do next but that it may not be able to provide a causal explanation of my behavior—why I did this rather than that, or what I meant when I said that my choice was voluntary. For that, we need to resort to high-level constructs, such as alternatives, reasons, and decisions. Moreover, the high-level constructs may be "multiply realizable" and thus not reducible to the neural substrate—another person may go through the same alternatives, reasons, and decisions, but these may be microphysically determined by different neural states.[xxxvi]

However, despite attempts to reconcile the microphysical substrate with high-level constructs, the extrinsic substrate$^{+}$ view remains wedded to the assumption that the microphysical substrate exists as such and that it is ontologically fundamental

---

[xxxiv] These conclusions do not depend on the size or complexity of the system to be simulated or on the size of the computer simulating it. (We are leaving aside questions related to fully neuromorphic computers and quantum computers.)

[xxxv] The substrate$^{+}$ view of seemingly "good enough free will" may well be compatible with determinism ((Dennett, 1984)), but that is beside the point here.

[xxxvi] Similar considerations apply not just to humans but also to complex artifacts endowed with autonomy and goal-directed behaviors. It does not make sense to try to account for the decisions made by an intelligent, autonomous robot purely based on the interactions among its transistors or atoms. Instead, we must resort to the high-level program that is guiding its actions. Moreover, the same behaviors can be implemented on different hardware platforms.

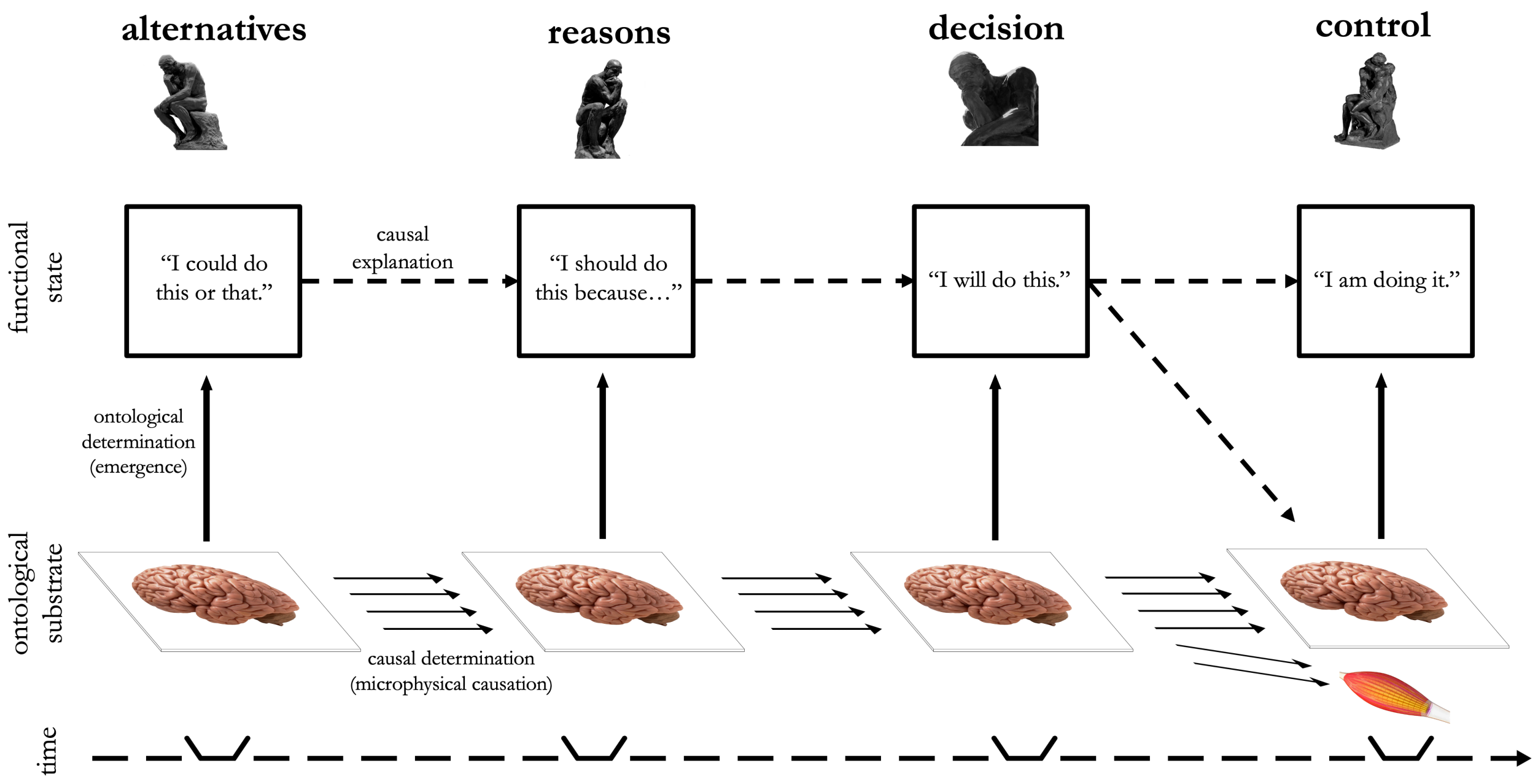

**Fig. 8** The extrinsic substrate+ view: no room for genuine free will. My experience of alternatives, reasons, a decision, and the feeling of control are functional states and constructs ontologically determined by the physical substrate of my brain (vertical arrows indicating "emergence"). Such constructs may be necessary for causal explanations across functional states (dotted horizontal arrows) and may be multiply realizable. Yet the true cause of each subsequent state is ultimately determined at the microphysical level (solid horizontal arrows).

or primary. High-level functions and constructs, as well as phenomenal states (to the extent that they are acknowledged), are ontologically derived or secondary because they are determined by the substrate and cannot exist without their substrate existing as such. The microphysical substrate is even more fundamental when it comes to causation. As we have seen, microphysical determination is assumed to be causally closed—it is sufficient to determine everything that happens next. This excludes higher-level causation: high-level properties and constructs cannot cause anything above and beyond what has already been caused at the microphysical level. Said otherwise, because of ontological ("vertical") micro determination, what I experience is determined by my neurons (which exist as such). And because of causal ("horizontal") micro determination, what I am going to do next is also determined by my neurons (which exist as such and are the exclusive physical cause of what happens next). Any notion of genuine or "metaphysical" free will is therefore an illusion: free will may be treated as existing in a practical and social sense, but it cannot possibly exist in a genuine sense.[xxxvii,xxxviii,xxxix] In the end, I always do what my neurons do—I am just "carried along for the ride."

By assuming the fundamental existence of microphysical substrates as such, the extrinsic substrate+ view inevitably leads to a cognitive dissonance. On one hand, I believe that my brain exists as such, and that it is made of neurons. I also believe that my feelings and thoughts emerge upon the brain's microphysical substrate. Finally, I know that neurons do what they do based on their inputs. Therefore, whatever my feelings and thoughts might be, I cannot help doing exactly what my neurons do. And because neurons don't have a choice, neither can I. On the other hand, I vividly experience alternative possibilities and struggle to make the right choice, and I cannot help feeling that I am the one deciding what to do. So it must be that my choice, my freedom, and my will, are ultimately illusory. As stated by Dr. Johnson, "all theory is against the freedom of the will; all experience is for it."

This dissonance is resolved only by abandoning the extrinsic substrate+ view and recognizing the intrinsic powers view as the proper ontological starting point (**tables III and IV**). To understand consciousness and free will, I must acknowledge that what exists absolutely is only my experience. I must envision that it can be accounted for, in physical terms, by the

[xxxvii] It is often pointed out that several empirical results confirm the illusory nature of free will. In Wegner-style experiments ((Wegner, 2002)), which manipulate delays and what the subject sees, a subject can be convinced that they were the agent moving their hand "of their own free will" when it was actually the experimenter who did so. Similar misattribution of agency can also occur during seances. These illusions demonstrate that, under special conditions, our experience of agency can be illusory but not that, under typical conditions of deliberation and action, our free will is an illusion.

[xxxviii] One way around this conclusion is to appeal to an immaterial soul that can suspend the workings of the brain, decide autonomously, and tell the neurons what to do. But barring such an interactionist-dualist leap of faith, there seems to be no way out.

[xxxix] In fact, in the extrinsic view, one might argue that genuine free will is simply an incoherent notion: determinism or not, we should "just get over it." This is because our actions are either fully determined by our neurons, in which case we have no freedom, or they are not, in which case they are due to chance, not to our will.

| **Table III: Intrinsic powers ontology** | **Table IV: Extrinsic substrate⁺ ontologies** |
|---|---|
| **Existence**<br>• Only intrinsic entities exist as such.<br>• A substrate is "physical" in an operational sense. It comes down to conditional probabilities among micro units that can be observed and manipulated, yielding a transition probability matrix (TPM).<br>• As such, it permits the operational determination of the $\Phi$-structure of an intrinsic entity by "unfolding" its cause– effect power (physical existence) based on the essential properties of experience (phenomenal existence).<br>• Alternatives, reasons, and decisions exist as sub-structures ($\Phi$-folds) within an intrinsic entity. | **Existence**<br>• Physical substrates exist as such.<br>• A substrate is "physical" in an ontological sense. It is constituted of micro units endowed with intrinsic physical properties (mass, charge, spin) and following the "laws of physics."<br>• As such, it ontologically determines or "gives rise" to emergent functions, such as decision-making, or emergent feelings, such as that of agency.<br>• Emergent functions or feelings are considered "real" insofar as they are necessary for explanation and because of their multiple realizability. |
| **Causation**<br>• Alternatives, reasons, and decisions genuinely exist, but the micro substrate from which they can be operationally unfolded does not exist as such.<br>• Decisions genuinely cause, because only what exists (intrinsically) can cause. Intrinsic causation is characterized by an $\mathcal{A}$-structure that links a preceding $\Phi$-structure with a subsequent $\Phi$-structure.<br>• Decisions exclude micro causes. | **Causation**<br>• Alternatives, reasons, and decisions are "carried along for the ride" by states of the microphysical substrate by which they are ontologically determined.<br>• Causation is microphysical. A substrate goes through a series of states, in which the present state of each micro unit is causally determined by inputs from its past state.<br>• Reasons and decisions do not cause but may be necessary for causal explanations and because of multiple realizability. |

$\Phi$-structure unfolded from my main complex. I must realize that the only alternatives, reasons, and decisions that can be properly said to exist are those that exist intrinsically, in my experience, corresponding to sub-structures within a $\Phi$-structure. It is thus wrong to think of what I am conscious of as being ontologically determined *by* my substrate, as if the substrate existed as such—as an aggregate of atoms that "gives rise" to my experience. Instead, the substrate must be considered as an operational basis *from which* one can operationally determine (unfold) the properties of experience. In short, I must realize that if my experience exists, then my neurons do not also exist, neither as a substrate as such nor as little entities of their own.

Similarly, I must realize that my decision is caused by my reasons and that it causes my action. Thus, it is wrong to think of my decision as being causally determined *by* the previous state of my substrate, as if that substrate existed physically as such and caused as such. Instead, the substrate must be considered as an operational basis *from which* one can determine (unroll or predict) the next state. But my decision is not caused by the substrate, because only what exists can cause, and what exists absolutely is only the $\Phi$-structure unfolded from a complex, not a substrate as such.

### 7.2 No choices and decisions, problems and solutions, questions and answers?

The intrinsic powers view of existence applies not only to alternatives, reasons, and decisions, but to everything that exists in the mind and thus exists intrinsically. For example, facing a choice and making a decision is similar in nature to many other high-level, abstract experiences, such as having a problem and finding a solution, or having a question and finding an answer. Compared to the intense debates about the possibility of genuine alternatives and genuine decisions, there has been little concern about whether there may be genuine problems and solutions, genuine questions and answers. The reason, most likely, is that the debate on free will has focused on determinism rather than on existence, and determinism may seem more problematic for achieving freedom than for finding answers or solutions.

But if we adopt the extrinsic substrate⁺ view, just as we cannot have a genuine choice and decide how to act, we cannot have a genuine problem and solve it, or have a genuine question and answer it. Like choices, problems and questions may be considered emergent constructs that are necessary for explaining our behavior and that cannot be reduced to a physical substrate owing to multiple realizability. But like decisions, solutions and answers must be determined by the microphysical substrate, which is assumed to exist as such and which inexorably updates its state, each one causing the next. Once again, we are just carried along for the ride, thinking we have solved the problem or answered the question.

In the intrinsic powers view, by contrast, there are genuine problems and genuine solutions, genuine questions and genuine answers. Like alternatives and decisions, they exist intrinsically as sub-structures within the $\Phi$-structure that corresponds to our current experience. In fact, problems and solutions, questions and answers exist *only* as contents of experience within individual subjects. They do not emerge from a neural substrate that exists in its own right, nor do they live in a separate realm of ideas. I have problems—not my neurons.

# 8 Freedom, indeterminism, and predictability

IIT's argument for genuine free will hinges on the proper understanding of experience as absolute existence and on the intrinsic powers view: what exists absolutely, in physical terms, are intrinsic entities, and only what exists can cause. On the other hand, much of the debate about free will has revolved not around existence but around the role of indeterminism. What does IIT have to say about indeterminism and its significance for free will?

We will briefly address three points. First, a degree of fundamental indeterminism is mandated by IIT's requirements for intrinsic existence. Second, this fundamental indeterminism is not the reason we have genuine free will; in the timeframe of a few seconds, our actions are likely highly determined and thus predictable. However, since existence is not constitution and causation is not prediction, genuine free will is not only compatible with a high degree of determinism but requires it. Third, fundamental indeterminism matters in the long term. If we have genuine free will, and if the future is not preordained, we are responsible not only for our present actions but, in a very real sense, for what they bring about in the future: our freely willed actions are both self-changing and world-changing.

## 8.1 IIT and fundamental indeterminism

IIT assumes an explanatory model based on cause–effect power "all the way down." To be self-consistent, this model requires that the "atomic" units of cause–effect power satisfy the requirements for intrinsic existence.[19] In principle, each unit in its current state must not only be able to take and make a difference from and to itself (*intrinsic specification*), but also be able to provide itself with a difference (*intrinsic differentiation*). In other words, it must both specify a cause and an effect state by increasing its probability and provide itself with an alternative of non-zero probability, regardless of the input from other units (the "background conditions"). This leads to a tension between intrinsic differentiation and specification. It also means that a degree of micro differentiation, which is to say of indeterminism, is required to ensure that intrinsic cause–effect power is available all the way down.

## 8.2 Free will and predictability

In a typical free will scenario, say, one taking place over a few seconds, micro indeterminism, say of the quantum kind, is unlikely to influence my decisions (and even if it did, it will not buy me any freedom). As a consequence, my actions should be largely predictable, at least in principle, say if we had a low-level, detailed model of my brain. In fact, they may often be partially and coarsely predictable even at a high, psychological level. For example, people who know me well may be able to correctly predict how I will choose in specific circumstances simply because they have a good understanding of my values and beliefs and how they might determine my actions. The question is whether predictability is incompatible with free will. For example, consider a scenario in which I am asked to choose to which charities I wish to donate (Maoz et al., 2019). One day, a computer simulating the interactions among my neurons with great precision should be able to achieve near-perfect prediction over the few seconds of my deliberations, as long as they take place in a controlled environment. Moreover, a very fast computer should be able to systematically predict my choices before I make them (Maoz, 2012). This would certainly evoke an eerie feeling of alienation—of being a mere "puppet." If every detail of my behavior is fully predictable in advance, how can I possibly be free?

As before, determination and the ensuing predictability seem to leave no way out for genuine free will. And, as before, the problem is that the notions of determination and predictability are based on the extrinsic substrate$^{+}$ view. In that view, my brain exists as such: it is a complicated physical substrate—a biological machine—that follows a single, predetermined trajectory. It has no alternatives or choice and, at least in the short run, it cannot help but "go through the motions." And while it may well be that my functional states, as well as my feeling of having alternatives and making decisions, cannot be reduced to their physical substrate in the brain, they are ontologically determined by it and are the result of microphysical causation. Inexorably, regardless of my protestations, I am just carried along for the ride. Because microphysical causation is what causes what happens next, there is no real choice, and predictability is incompatible with free will (nor would unpredictability).

But if we start from phenomenology and take the intrinsic powers view, the perspective on what actually exists and causes changes radically. My freedom depends on the intrinsic existence of choice—of multiple alternatives in my mind, here and now—not on the possibility of multiple trajectories unrolling over time. Accordingly, *it is not chance but choice that makes us free—not a fork in the road ahead (−<) but a fork in the mind here and now ( Y ).* Trajectories themselves are just useful predictive devices for conscious observers (equipped with powerful computers): they do not exist as such, but only in the mind of the observer who can interpret them.[xl]

[xl] The availability of alternative extrinsic trajectories due to indeterminism is not necessary for freedom, but neither is it sufficient. If I am not conceiving alternatives in my mind, it is irrelevant that indeterminism in my brain might leave open multiple trajectories for my future actions. Whatever action occurs would be due to chance and not to will.

In summary, high-level predictability is not a problem for free will: if my alternatives are evaluated based on my reasons, my reasons cause my decisions, and my decisions cause my actions, no wonder prediction is possible. If I have strong reasons for choosing the way I did, then under the same circumstances, I should freely make the same choice, again and again. Nor is low-level, microphysical predictability a problem. A detailed, low-level model of the main complex is one that offers a good operational account of what exists intrinsically (by unfolding $\Phi$-structures) and of what causes intrinsically (by computing $\mathcal{A}$-structures originating in $\Phi$-structures). If the model of what exists and causes is good, it will also predict well what will exist next (my decision will come into being, caused by my reasons, and it will cause my action). But if the model is good, it will also offer a shortcut to predict what happens next without even bothering to unfold what exists and causes: we simply need to unroll the successive states of every unit based on its inputs and mechanism.[xli] In short, genuine free will is fully compatible with full predictability.

From this perspective, it is justified to have a feeling of alienation if confronted with a computer simulation predicting my decision: the simulated neurons are indeed "not me"—they don't exist or cause, even though they can be a convenient operational shortcut for unfolding what exists and causes and predict what happens next.

### 8.3 Historical determination and personal responsibility

We have seen that short-term determinism, whether at the level of my values or my neurons, is compatible with genuine free will and not just with free will of convenience.[xlii] Nevertheless, modern physics assumes that a degree of fundamental micro indeterminism at the quantum level is part of the fabric of nature. As described above, the same assumption is made by IIT purely as a requirement for intrinsic existence. With respect to free will, the main consequence is not that fundamental indeterminism makes us free, but that it makes the unrolling of the universe unpredictable in principle, hence not preordained, at least in the long term. Therefore, given fundamental indeterminism, genuine free will, and considering the ability of freely willed actions to modify their own substrate, we are genuinely responsible not just for our actions here and now, but also for how they will change ourselves and our environment in the future. Conversely, if determinism were true, it would never have been causally possible for anything, including my actions or resolutions, to have gone any different than they did. Instead, indeterminism makes it open-ended, and allows us to steer the future one way or another.

To see why, the contrast of views is, again, helpful. In the extrinsic substrate+ view, the universe unrolls according to the "laws of physics," and under certain conditions, it may "give rise" to physical substrates of increasing complexity, such as those that can support life, multicellular organisms, and biological intelligence (**fig. 9**). Eventually, highly evolved animals such as ourselves become endowed with sophisticated emergent functions, including rational thought and language, which allow us to build societies and cultures. The evolution of the universe, as well as the precise course of evolutionary and cultural events on earth, are fundamentally unpredictable due to the inherent indeterminism of the microphysical world. As we have seen, however, indeterminism does not buy us freedom, just some measure of unpredictability. Everything that happens is either determined by the previous state of the universe, in which case we can say that it is caused at the microphysical level, or it is not determined, in which case we can say that it occurs by chance (or it is a mixture of determinism and chance). In all cases, my emergent functional (and phenomenal) states can only "go along for the ride," without exerting any real influence on what happens next.

In such a scenario, nothing is genuinely my responsibility; everything is either determined by previous occurrences or just happens by chance. In the extrinsic substrate+ view, then, I can only be held responsible in a pragmatic, socially motivated sense. Accordingly, an enlightened society may decide to employ preventive dissuasion rather than punishment, but in the end, even that societal decision is not genuinely free.[xliii]

The intrinsic powers view is radically different (**fig. 10**). At some point in the history of the earth, organisms evolved that harbored neural substrates unfolding into $\Phi$-structures of high $\Phi$.[xliv] These organisms have full-fledged experiences: there is much "it is like to be" them because they exist as large intrinsic entities. Consciousness may grow, partly favored by selective pressure towards integration. However, as we have also seen, consciousness alone is not sufficient for free will. For that, organisms must evolve who are able to consciously envision alternatives, assess them against reasons, decide on the basis of those reasons, and intend, cause, and control actions. Conscious notions of the self, of values, and of beliefs are also typically needed, and so is conscious understanding of the context of one's choice. When all these requirements are met, free will appears and, with it, *responsibility*.

Once free will comes into being on the face of the earth, it

---

[xli] If the model is correct, the prediction will be correct. But that does not mean that the model exists or causes. Imagine the model were simulating the cerebellum or another computer; the prediction would be just as good, but neither the model nor what it models would exist or cause. More generally, prediction is blind to existence, borders, grain, and structure ((Grasso et al., 2021)).

[xlii] This refers to the pragmatic variety of free will used by the legal system and throughout society, which holds that adults are responsible for their actions except under conditions of duress, insanity, and so on.

[xliii] This sentiment was expressed well by Omar Khayyam: "And that inverted Bowl we call the Sky, Whereunder crawling coop't we live and die, Lift not thy hand to It for help—for It Rolls impotently on as Thou or I."

[xliv] Studies using simulated organisms with a simple brain, called *animats*, illustrate how selective pressure may favor the evolution of substrates having high $\Phi$ ((Albantakis et al., 2014)).

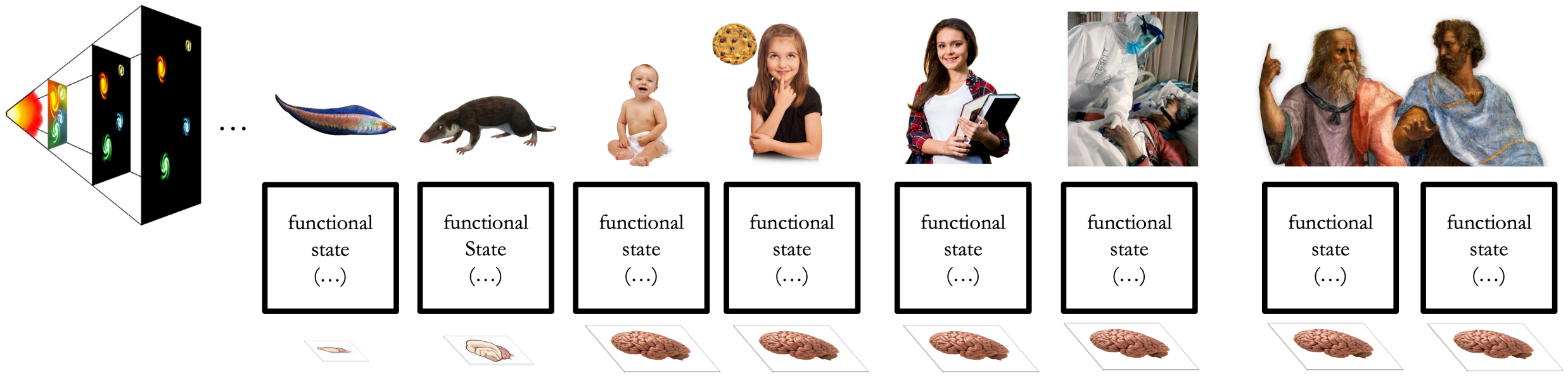

**Fig. 9** Historical determination and personal responsibility: the extrinsic substrate⁺ view. Governed by the "laws of physics," microphysical substrates give rise to basic life forms, which eventually evolve into animals, such as myself, with sophisticated emergent functions, such as decision-making, language, and rationality. However, these functions emerge upon the microphysical substrate and are "carried along for the ride." Hence, nothing is truly my responsibility but is determined by previous occurrences and chance at the microphysical level.

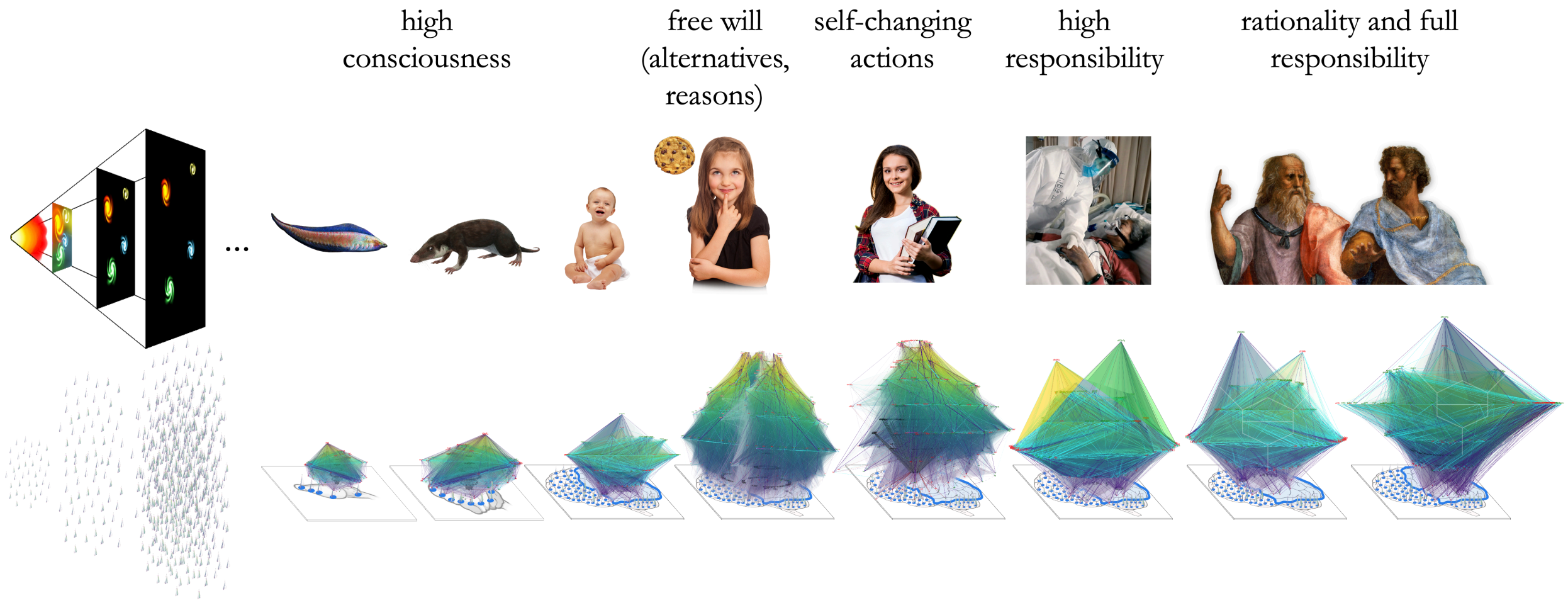

**Fig. 10** Historical determination and personal responsibility: the intrinsic powers view of IIT. As the universe evolves, what first exists are merely small aggregates of ontological "dust"—intrinsic entities characterized by negligibly small Φ-structures. Organisms eventually evolve with neural substrates, portions of which unfold into Φ-structures of high Φ. Genuine free will becomes possible in an intrinsic entity, such as myself, with a Φ-structure complex enough to conceive of alternatives—a "fork in the mind"—and to reason about them. My capacity for genuine free will grows the more I can exercise self-changing actions—those that modify my own substrate—along with world-changing actions—those that modify my surroundings. These capacities enable me to develop rationality and render me more and more responsible for my decisions. As such, with full rationality comes full responsibility.

begins to transform those who harbor it and the world that surrounds them, molding both self and environment towards values and purposes. Freely willed *self-changing actions*[xlv] are those that cause a long-lasting change in the mechanisms and connectivity of my substrate—that is, ultimately, in its TPM.[xl] For example, suppose I decide to devote myself to the study of medicine to improve the human condition. Over time, what I study will change the connections among neurons in my substrate and thereby change who I am—my reasons, values, beliefs, and even the strength of my resolve (or willpower). Or I may decide to devote myself to the preservation of the environment, changing not just myself but also what surrounds me. My actions then become *world-changing actions*.

The more I perform self-changing and world-changing actions out of my own free will, the more I acquire genuine *responsibility*. If I have free will, my freely willed actions cause what will happen in the long run, what I will become, and how I will affect the world, molding myself and environment according to my purpose, not at the whim of chance. Furthermore, my personal responsibility can only grow with increased consciousness and understanding—that is, with increased freedom of my will. Finally, if I cultivate the ability to assess and criticize reasons, values, and beliefs in a *rational* manner, I will become fully free and thereby fully responsible. This is because my choices will no longer be constrained by my own subjective beliefs alone, but they will be evaluated under the objective (and inter-subjective) light of reason. With freedom, indeed, comes responsibility, and with that the

[xlv] The term is similar to Kane's "self-forming actions" ((Kane, 2002)). However, in IIT self- changing actions are free not through indeterminism, but because they are caused by an intrinsic entity who conceives alternatives and reasons, makes decisions, and controls actions.

opportunity to shape an open-ended future according to our will, guided by rationality.

These conclusions derived from the intrinsic powers ontology and its account of causation happen to align with both common sense and many spiritual, religious, and wisdom traditions: we can be the authors of our deliberate actions, and we bear responsibility for their consequences. As a popular saying goes, "Watch your thoughts, they become your words; watch your words, they become your actions; watch your actions, they become your habits; watch your habits, they become your character; watch your character, it becomes your destiny."[xlvi] Conversely, as discussed in **section 6.3**, digital computers as currently built support no significant $\Phi$-structure, no matter how intelligently they act. They do not exist for themselves, intrinsically, but only for us as conscious observers. Thus, according to the intrinsic powers view, they cannot genuinely cause nor are they responsible for anything. This has important legal implications for how we build and operate them.

# 9 IIT's view of free will in relation to major philosophical positions

Having presented IIT's implications for free will, it may be useful to consider how IIT's account compares to some prominent positions in the philosophical and neuroscientific debate. In philosophy, free will is commonly understood as requiring two conditions: alternative possibilities and self-determination. The former refers to the ability to do otherwise—that is, that an agent could have acted differently had they willed so. The latter demands that the agent is the source of her own actions and that these actions are determined autonomously, for example, by the agent's beliefs and intentions. In both the historical and contemporary debate, many have taken a stance on the existence of free will and on whether or not it is compatible with determinism.

## 9.1 Denying free will

A classic position argues that determinism excludes both the existence of alternative possibilities and the agents' ability to self-determine. In the 18th century, philosophers such as Baron d'Holbach and Julien Offray de La Mettrie argued that human beings are entirely governed by physical laws, just like any other part of nature (d'Holbach, 1770; La Mettrie, 1747). Since we are biological machines, there is no room for metaphysical free will, and our sense of freedom arises from our ignorance of the causal factors shaping our decisions.

A contemporary rendition of the same position, grounded in biological evidence, is found for example in Robert Sapolsky's view that every human action is fully determined by prior genetic and environmental factors (Sapolsky, 2023). From a scientific perspective, humans are complex biological machines whose behaviors could in principle be predicted given enough information. The intuition that we have alternative possibilities—that we could have acted otherwise—is simply a cognitive illusion. Consequently, holding people morally responsible is no more coherent than blaming natural phenomena for causing disasters.

Sam Harris also argues that our choices are determined by prior causes, including genetic, environmental, and neural factors (Harris, 2012). He points out that thoughts and intentions appear in our minds spontaneously, without our direct control or authorship, and this is true for decisions too. Drawing on neuroscientific evidence such as Libet's experiments, he emphasizes that decisions are initiated in the brain prior to our awareness of making them. Since we do not consciously control the origin of our thoughts, and given the temporal lag between neural activity and conscious intention, Harris concludes that our sense of agency is merely an illusion, with no metaphysical grounding.

As already pointed out by Cicero, Gassendi, David Hume, and more recently Roderick Chisholm (Chisholm, 1966), indeterminism is no better ground for the possibility of metaphysical free will: chance does not buy us any freedom because randomness and unpredictability can only hinder the agent's capacity to self-determine. This has led philosophers such as Galen Strawson and Derk Pereboom to argue that, since free will is incompatible with both determinism and indeterminism, it is an incoherent notion (Pereboom, 2009, 2014; Strawson, 1994).

In general, these positions are in line with the *substrate only* ontologies discussed above. These views take the physical substrate—either neurons or the molecules and atoms constituting them—as the fundamental level of reality. Consciousness, together with the reasons, beliefs, and decisions of the agent, are seen as cognitive illusions without causal power of their own. From this perspective, persons do not exist as agents, and do not cause; instead, they are epiphenomenal byproducts of neural processes.

## 9.2 Trying to preserve free will by redefining it

Others have insisted that the concept free will cannot be dismissed entirely. Without it, we would lack the conceptual tools to distinguish between actions attributable to an agent and those resulting from coercion or constraint—a distinction that plays a key role in legal settings, where it matters for the assignment of responsibility and punishment. In response, several positions have sought to redefine free will by relaxing its metaphysical commitments.

A common position is to view free will as compatible with determinism, regardless of whether the universe is fully deterministic or not. Classical compatibilists from the 17th century such as Thomas Hobbes and David Hume defined freedom as the ability to act in accordance with one's desires,

[xlvi] This quote has been credited to various historical figures, including Lao Tzu, Ralph Waldo Emerson, and Gautama Buddha. While such figures have made similar statements, the true source appears to be less exotic: Frank Outlaw, founder of BI-LO supermarkets.

independent of external constraints (Hobbes, 1651; Hume, 1748). Hume maintained that human liberty requires that our actions flow from our internal motivations rather than external coercion. The ability to do otherwise is interpreted in conditional terms: a person could have acted differently if they had different desires or motives.

Contemporary compatibilists refine this idea by focusing on authenticity and rational responsiveness. Harry Frankfurt introduced the hierarchical model of free will to capture the distinction between actions that are merely caused and those that are genuinely controlled by the agent (Frankfurt, 1969). He distinguishes between first-order desires, which motivate immediate actions and may reflect environmental triggers or compulsions, and second-order desires, which express what the agent wants their motivations to be. According to Frankfurt, an agent acts freely when their effective first-order desire is endorsed at the level of second-order volition—that is, when it reflects what the agent identifies with.

Another compatibilist position, developed by John Martin Fischer and Mark Ravizza, emphasizes reasons-responsiveness, a revised notion of alternative possibilities (Fischer & Ravizza, 1998). The agent's freedom depends not on the availability of metaphysically open options, but on the capacity to respond appropriately to reasons, meaning that if a compelling reason to act otherwise had been present, they would have acted differently. R. Jay Wallace advances a similar position, highlighting the role of reflective self-control in moral responsibility (Wallace, 1994). According to this perspective, determinism does not undermine responsibility so long as individuals possess the capacity to regulate their behavior in accordance with rational considerations.

A defense of free will as compatible with determinism has also been provided by Daniel Dennett, who redefines its ontological status as an evolutionarily advantageous "user illusion:" a practical construct that helps humans function in complex social environments (Dennett, 1984; Dennett, 2003). By adopting the intentional stance—treating others as agents with beliefs and desires—we make sense of behavior and enable practices like moral responsibility. Dennett's view redefines free will in instrumental terms: this pragmatic notion of freedom is all we need, and all we could rationally want, even if it is ultimately illusory, lacking a strong metaphysical status.

More recently, Christian List has offered a view that aims to preserve genuine agency as an emergent property tied to the unpredictability of the agent's choices at the macro level (List, 2019). Even under determinism, argues List, the same macro level state—corresponding to, say, wondering to which charity I should donate—is realized by many micro states. Different micro states may determine different macro outcomes, so when I make my choice, it is genuinely unpredictable for me as an agent.

A different compatibilist argument ties the existence of free will to unpredictability. Seth Lloyd argues that even if the universe operates deterministically, it cannot fully predict its own future (Lloyd, 2012). This is because any attempt at complete self-prediction would require the universe to simulate itself in perfect detail, which leads to logical limitations akin to Gödel's incompleteness theorems and Turing's halting problem. As a result, the future remains fundamentally unknowable.

Most of the positions discussed in this section implicitly or explicitly embrace substrate$^+$ or multilevel ontologies. On these views, the physical world at the micro level may well be largely deterministic and causation may proceed micro state by micro state, but agents and their decisions are granted emerging existence at a higher level, where their actions can only be explained in terms of mental contents such as reasons and intentions. In this framework, free will is preserved not by rejecting micro causation, but by situating freedom at a different level of description—the macro level of agents. Clearly, such redefinitions will not satisfy those who deny fundamental free will, because the impossibility of objective self-prediction provides freedom from predictability, not freedom of choice, and subjective ignorance can only provide the illusion of free will.

### 9.3 Trying to save free will, but with the wrong medicine

By contrast, libertarian positions assert the existence of metaphysical free will and ground it in indeterminism. Libertarians argue that genuine freedom requires indeterminacy in the decision-making process. One prominent libertarian model has been developed by Robert Kane (Kane, 1996, 2002). Kane emphasized critical situations of moral conflict where the agent must make an effort to decide among courses of action that are equally motivated. Which course is chosen is not determined—indeterminism intervenes like a micro coin flip—but the action is far from not been caused. Kane also emphasized that these critical junctures lead to self-forming actions, which serve as the foundation for the agent's long-term character and bestow responsibility upon future actions. However, Kane's recourse to indeterminism to flip the coin cannot avoid the criticism that the available courses of action can still be considered a necessary product of past occurrences, with some chance superimposed.

A more radical libertarian position is agent causation, the view that a free action is caused not by prior events but by the agent as a substance or self, who initiates new causal chains independently of deterministic or random processes. Roderick Chisholm and Timothy O'Connor defend this view, arguing that a person, as a persisting substance, exercises a special causal power (agent causation) that is irreducible to the physical causation at play in the natural world (event causation) (Chisholm, 1966; O'Connor, 2000). In this framework, agents act as "prime movers unmoved," initiating new causal chains rather than being determined by prior events. This approach aims to secure ultimate self-determination, but at the cost of metaphysical plausibility, particularly regarding how such agent causation fits within our scientific understanding of the world.

Some libertarians, including Carl Ginet and Hugh McCann,

propose instead non-causal accounts of free will, suggesting that volitional acts do not require causal determination at all (Ginet, 1990; McCann, 1998). However, if free actions lack causal ties to an agent's reasons or character, they risk being arbitrary and unintelligible. Without such grounding, actions may not reflect the agent in a morally significant way, undermining both agency and responsibility (Clarke, 2003; Frankfurt, 1969).

### 9.4 IIT and the philosophy of free will

Compared to the philosophical positions on free will reviewed above, IIT neither denies its existence, nor attempts to rescue it by redefining it, nor seeks to preserve it through a special kind of causation. Instead, it grounds free will in what exists and what causes.

Most views that deny the existence of free will assume a substrate-only ontology: that the physical substrate, whether neurons or atoms, exists as such and constitutes the fundamental level of reality. From this perspective, consciousness, along with alternatives, reasons, and decisions do not exist—not in a fundamental sense—and agents do not genuinely cause—they are merely aggregates whose causal powers are exhausted at the micro level. In contrast, IIT begins with the experience itself and rejects the assumption that the substrate exists as such. As emphasized above, existence is not constitution: a substrate is only an operational basis from which cause–effect power can be unfolded to reveal what exists intrinsically. Agents, as intrinsic entities, exist absolutely and cause intrinsically. If I contemplate alternatives and decide among them, those alternatives exist as intrinsic sub-structures within my experience and my decision is genuinely free—it is I who cause, not my neurons.

Other views attempt to preserve free will by redefining it in terms of reasons-responsiveness, self-control, hierarchical desires, or unpredictability at the macro level. These views all implicitly assume a substrate$^{+}$, multi-level ontology: they still take the microphysical substrate to exist as such, but treat agency, intentions, and reasons as emergent properties at a higher level of description. In this framework, higher-level constructs are granted existence because they are required for explanation and they are multiply realizable, so they cannot be reduced to the micro level. Therefore, freedom would remain necessary as an explanatory construct, and so would agent causation, even though genuine causation only happens at the micro level. As we have seen, IIT only recognizes absolute existence to intrinsic entities. Intrinsic entities with rich $\Phi$-structures are meaningfully conscious and, in our case, can be conscious of alternatives, values, reasons, and goals. These do not emerge from a substrate, but exist intrinsically, here and now, as cause–effect sub-structures within an experience. And only intrinsic entities, existing absolutely, can genuinely cause. IIT grants extrinsic existence to other levels of organization, which are important for explanation, but their existence is relative to us as conscious observers, not absolute.

Finally, IIT may seem aligned with libertarianism in affirming the existence of alternative possibilities, of an open future due to micro indeterminism, and of the genuine ability of agents to cause their own actions. However, IIT departs from it because of its intrinsic ontology. It construes alternatives as a fork in the mind rather than a fork in the road—a structure within experience rather than a branching trajectory in time. It grants a role to fundamental indeterminism ("differentiation" in IIT) not to ensure freedom of choice, but only freedom from predictability. And, because only what exists absolutely can cause, IIT does not require a special kind of uncaused agency: intrinsic causation suffices.

## 10 Some experimental tests

This essay ends with a few considerations about empirical tests. Scientists typically take the extrinsic substrate$^{+}$ view (at least implicitly) and tend to see the brain as a complicated biological machine. Like any machine—say, a computer executing a program—the brain cannot help but "run through the motions." In fact, neurophysiological findings are often presented as supporting the conclusion that free will is illusory (Libet et al., 1983; Soon et al., 2013). For example, a subject's decision, such as "I will move my arm," may be predicted above chance well before the subject "becomes aware of that decision" by recording correlates of neural activity in certain parts of the brain. And indeed, by investigating the brain as a physical substrate that exists and causes as such, it is inevitable that we will only find a series of neuronal events that predict other neuronal events, unrolling along tracks set by neuronal connections. Even if we accept the reality of consciousness and of feelings of free will, we must also accept that they emerge upon those neuronal events, that they may emerge late, and that, at any rate, they can play no role in what actually happens. At best, we may find some brain regions that are systematically activated or deactivated when a choice is entertained, a decision is made, and an action is experienced as willed (Haggard, 2019). But we will find nothing that could make any difference above and beyond what neurons have caused.

If we take IIT's intrinsic powers view, however, the perspective changes, and not just as a matter of interpretation. Rather, the validity of the intrinsic powers view can be tested objectively by assessing several specific predictions. For example, consider four paradigmatic scenarios in which the same outward action is performed (**fig. 11**). In the first condition, a subject performs either a left- or a right-hand movement as a reflex triggered by a percussion of the bicipital tendon. In the second, the subject does so due to transcranial magnetic stimulation (TMS) of the motor cortex. In the third, the subject does so voluntarily whenever they "feel the urge," in a Libet-style paradigm (Libet et al., 1983; Soon et al., 2013). And in the fourth, the subject does so voluntarily after deliberating which non-profit foundation should receive their donation (Maoz et al., 2019).

These four scenarios can be investigated in humans using functional MRI, high-density EEG, and TMS, as well as intracranial recordings and stimulations. As we have seen, based on the principles of IIT, the main complex should correspond to a maximum of irreducible cause–effect power

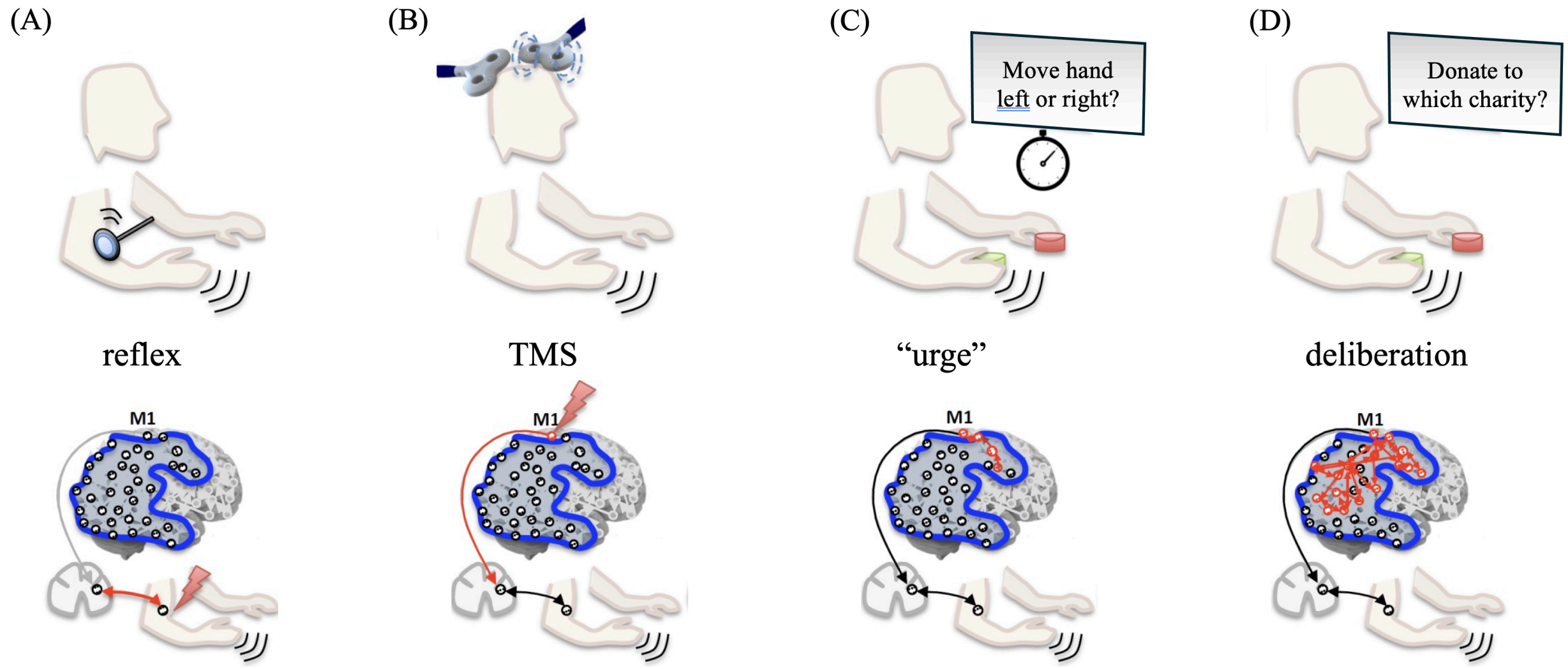

**Fig. 9** Human experiments: involuntary vs. voluntary actions, small vs. big decisions. The four panels show paradigmatic scenarios in which the same outward action (hand movement) is performed, but triggered differently. In each case, the neuronal sequence leading to the movement (shown in red) could be recorded using fMRI, high-density EEG, or intracranial recordings. A) The hand moves as a reflex triggered by an automated percussion of the bicipital tendon. IIT predicts that causal analysis should not trace the cause back to the main Φ-structure (blue outline). B) The hand moves due to a direct transcranial magnetic stimulation of the motor cortex ("M1"). Again, the cause should not originate within the main Φ-structure. C) The subject voluntarily moves their hand whenever they "feel the urge" in a Libet-style paradigm. The cause should originate within the main Φ-structure, albeit from a small sub-structure and with low $\mathcal{A}$—corresponding to a free but trivial decision. D) The subject voluntarily moves their hand after pondering which non-profit foundation should receive their donation. The cause should originate from a large sub-structure within the main Φ-structure and have high $\mathcal{A}$—corresponding to a free non-trivial decision.

($\varphi_s$) and its unfolded cause–effect structure should correspond to a subject's experience. Ongoing work using approximate measures suggests that it is possible to identify such a maximum of $\varphi_s$, primarily over posterior-central cortical regions. Encouragingly, the same set of posterior-central regions appear to be critical for consciousness and its various contents based on independent sources of evidence—including lesion, stimulation, and recording studies.[22] On this basis, it becomes possible to investigate several predictions of IIT that specifically concern free will. For example:

1) Do voluntary actions, but not reflex actions, originate within the main complex in the brain?
2) Is the $\mathcal{A}$ value higher, and the associated sub-structure within that $\Phi$-structure larger, for deliberate actions vs. actions triggered by urges?

In this way, propositions that may at first seem hopelessly "metaphysical"—about the nature of consciousness and free will—become fully physical (operational) and therefore testable.

# 11 Conclusion

The argument for genuine free will presented here depends on IIT as a theory of consciousness, the validity of which must first be evaluated at the conceptual level. Do the axioms fully capture the essential properties of consciousness—those that are immediately and irrefutably true of every conceivable experience? Do the postulates adequately formulate these axioms into physical, operational propositions? Can the explanatory identity account, at least in principle, for both the essential and the accidental properties of every conceivable experience?

Second, the theory must be evaluated empirically. Does consciousness go along with a maximum of $\varphi_s$ associated with a high value of structure integrated information ($\Phi$)? Does that maximum disintegrate when consciousness is lost? Does the empirical substrate of consciousness in our brain correspond to such a maximum—the main complex? Do the units of the main complex in our brain have a grain that maximizes $\varphi_s$? Can the cause–effect sub-structures unfolded from portions of the main complex account for basic aspects of experience, such as spatial extendedness, temporal flow, objects, and qualities such as colors and sounds?

If IIT survives these tests, then its implications for free will become relevant. What exists and causes are only intrinsic entities—absolute maxima of cause–effect power, corresponding to conscious beings such as ourselves. Further, if we can consciously formulate alternatives, consider reasons, make decisions, and control our actions, then we have genuine free will. The amount of free will for any decision can also be measured, and the origin of the decision can be established empirically, thereby providing further tests of IIT.[xlvii] Finally, if we do have free will, we can perform self-changing and world-changing actions that shape ourselves and our environment according to our intentions. And given the fundamental indeterminism that makes the world open ended in the long run, we can freely shape how we and the world will evolve. For this, we have ultimate responsibility.

**Competing interests**: G.T. and C.K. are Board Members and have a financial interest in Intrinsic Powers Inc. G.T. holds patent rights along with the Wisconsin Alumni Research Foundation: patent No. US 8,457,731 B2 and, with C.K., patent applications Wisconsin Alumni Research Foundation No. P220191, P220192.

**Funding**: This project was made possible through support from Templeton World Charity Foundation (nos. TWCF0216 and TWCF0526). The opinions expressed in this publication are those of the authors and do not necessarily reflect the views of Templeton World Charity Foundation.

**Author contributions**: Conceptualization: GT; Writing – original draft: GT; Writing – review and editing: GT, LA, MB, CC, MG, CK; Visualization: GT, MG; Funding acquisition: GT; Investigation: GT, LA, MB, MG, CK; Methodology: GT, LA.

**Acknowledgements**: We thank F. Ellia, G. Findlay, A. Haun, B. Juel, W. Marshall, W. Mayner, and especially J. Hendren and J.P. Lang for their comments on the manuscript and help with the figures.

[xlvii] Tests in humans are critical, because only in ourselves do we have direct evidence for experience, including the experience of free will. Experiments in non-human primates or rodents may also help the interpretation of human experiments. A complementary approach consists in testing the self-consistency of the IIT account of consciousness and free will using simulated agents (*animats*) ((Albantakis et al., 2014)). With such agents, the four scenarios described in humans can be reproduced in a simplified manner to systematically evaluate cause–effect structures and causal accounts. Specifically, we can 1) fully unfold $\Phi$-structures and measure their $\Phi$ value, 2) determine at which spatio-temporal grain the agent exerts a maximum of causal power, 3) back- and forward-track causes and effects while considering their optimal grain, 4) precisely establish whether the distal cause of an action originates within the agent's substrate, and 5) obtain a causal account and estimate $\mathcal{A}$-structures and their $\mathcal{A}$ value to measure how much the agent causes an action—in other words, to measure the amount of free will it exerts

# Supplementary information

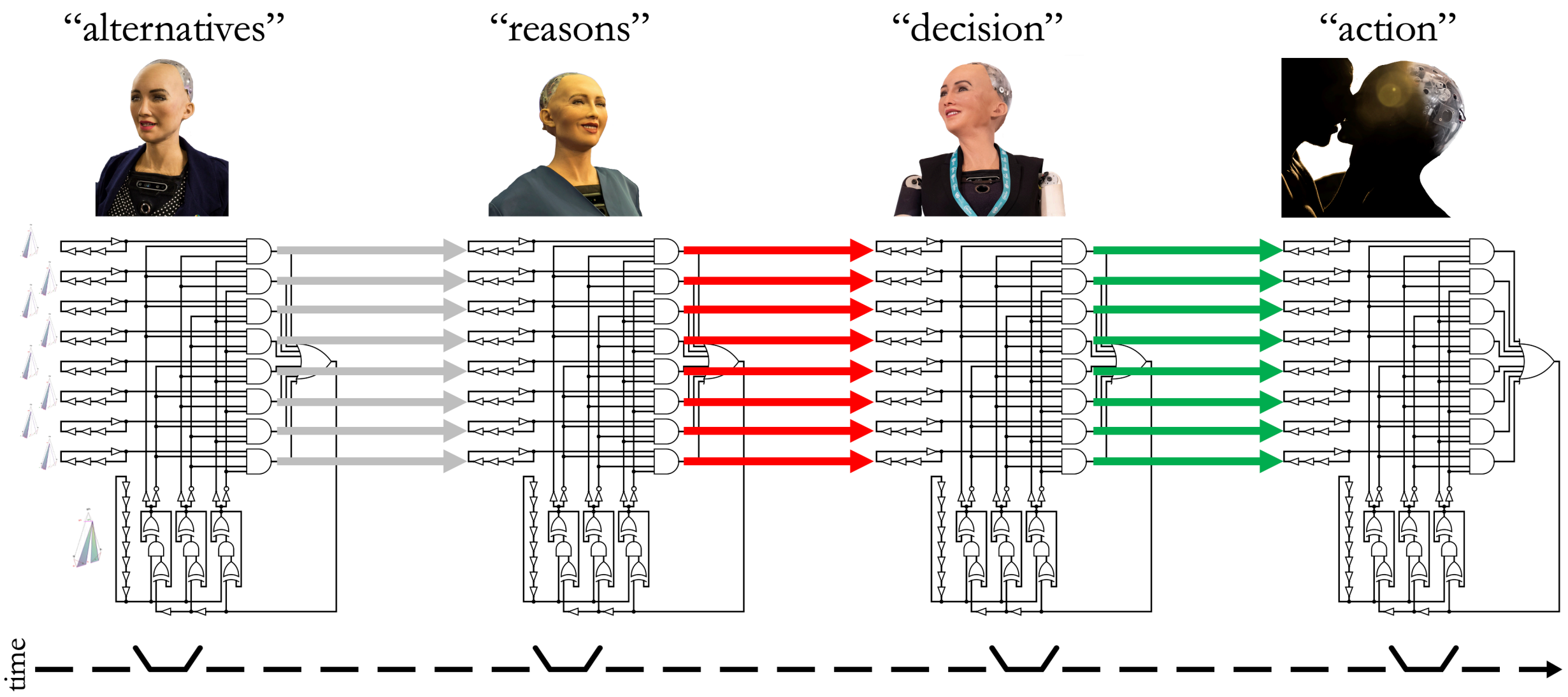

**Fig. 10** The “unrolled” causal structure of a computer simulating a brain. If one were to apply IIT’s ontological and causal analysis to a computer simulating a cognitive function—say, making a “decision,”—one would find that the computer as a whole does not exist as an intrinsic entity. Instead, its causal powers unfold into ontological “dust”—many minuscule entities (specified, say, by small sets of transistors arranged in a cycle, which unfold into tiny $\Phi$-structures [far left]). Moreover, the causal account would fragment into causal “dust”—many parallel causes and effects of very low causal strength $\mathcal{A}$ (shown in red arrows on the cause side and green arrows on the effect side).